\documentclass[11pt,a4paper]{article}
\pdfoutput=1

\usepackage{bm}
\usepackage{amsmath,amssymb}
\usepackage{cite}

\usepackage{graphicx}
\usepackage{color}

\usepackage{hyperref} 

\numberwithin{figure}{section}
\numberwithin{equation}{section}

\newcommand{\be}{\begin{equation}}
\newcommand{\ee}{\end{equation}}
\newcommand{\bea}{\begin{eqnarray}}
\newcommand{\eea}{\end{eqnarray}}
\newcommand{\vp}{\varphi}
\newcommand{\bp}{\bar \varphi}
\newcommand{\pb}{\bar \varphi}

\newcommand{\al}{\alpha}
\newcommand{\Vh}{\hat V}
\newcommand{\Kh}{\hat K}
\newcommand{\ph}{\hat \phi}
\newcommand{\kh}{\hat k}
\newcommand{\gmn}{g_{\mu\nu}}
\newcommand{\tgmn}{\tilde g_{\mu\nu}}
\newcommand{\bgmn}{\bar g_{\mu\nu}}
\newcommand{\hmn}{h_{\mu\nu}}
\newcommand{\thmn}{\tilde h_{\mu\nu}}
\newcommand{\hess}{\Gamma^{(2)}}
\newcommand{\eipx}{\,{\rm e}^{i p\cdot x}}
\newcommand{\emipx}{\,{\rm e}^{-ip \cdot x}}
\newcommand{\ex}[1]{\,{\rm e}^{#1}}
\newcommand{\dclnf}{\,\partial_\chi\! \ln\! f \,}
\newcommand{\cb}{\bar \chi}

\newcommand\ie{\textit{i.e.}\ }
\newcommand\eg{\textit{e.g.}\ }
\newcommand\cf{\textit{cf.}\ }

\newcommand{\etc}{{\it etc.}\ }
\newcommand{\viz}{{\it viz.}\ }
\newcommand{\half}{\tfrac{1}{2}}

\newcommand{\eps}{\varepsilon}

\begin{document}
\begin{titlepage}

\begin{center}
{\huge \bf Background independent exact renormalization group for conformally reduced gravity}
\end{center}
\vskip1cm


\begin{center}
{\bf Juergen A. Dietz and Tim R. Morris}
\end{center}

\begin{center}
{\it School of Physics and Astronomy,  University of Southampton\\
Highfield, Southampton, SO17 1BJ, U.K.}\\
\vspace*{0.3cm}
{\tt  J.A.Dietz@soton.ac.uk,\\ T.R.Morris@soton.ac.uk}
\end{center}

\abstract{
Within the conformally reduced gravity model, where the metric is parametrised by a function $f(\phi)$ of the conformal factor $\phi$, we keep dependence on both the background and fluctuation fields, to local potential approximation and $\mathcal{O}(\partial^2)$  respectively, making no other approximation. Explicit appearances of the background metric are then dictated by realising a remnant diffeomorphism invariance. The standard non-perturbative Renormalization Group (RG) scale $k$ is inherently background dependent, which we show in general forbids the existence of RG fixed points with respect to $k$. By utilising transformations that follow from combining the flow equations with the modified split Ward identity, we uncover a unique background independent notion of RG scale, $\hat k$. The corresponding RG flow equations are then not only explicitly background independent along the entire RG flow but also explicitly independent of the form of $f$.  In general $f(\phi)$ is forced to be scale dependent and needs to be renormalised, but if this is avoided then $k$-fixed points are allowed and furthermore they coincide with $\hat k$-fixed points.}



\end{titlepage}
\tableofcontents

\section{Introduction}
\label{sec:Intro}
Since gravity and quantum mechanics must somehow live successfully together, a quantum field theory of the metric $g_{\mu\nu}$, or something akin to the metric, has to be a sensible language at least over many orders of magnitude in length. We quantify this through some length scale $1/k$ (where $k$ has units of momentum, mass or energy). But since length scales must themselves be fluctuating quantities in a quantum gravitational theory, it is not {\it a priori} clear what one should mean by the scale $k$ in this context.  This matters because in quantum field theory, quantities, and couplings in particular, run with scale. It appears as though the very notion of such a scale must remain  inherently dependent on the metric $g_{\mu\nu}$ itself. These issues become especially acute when we try to formulate the question non-perturbatively.\footnote{Actually the issue exists even if nothing runs with scale. Operator product expansions within conformal field theory then in principle avoid introducing a scale but it is unclear what these constructs mean in a fluctuating geometry.} Indeed we will demonstrate in this paper that at least in the conformally reduced gravity model, this dependence in general leads to a direct conflict with basic RG notions such as fixed points with respect to the scale $k$. 

As we will discuss in more detail shortly, the non-perturbative approach most often used, addresses these issues by using the background field method. Then the challenge becomes
recovering a truly background independent description starting from flows in the inherently background dependent scale $k$. At the same time, in functional RG approaches another challenge is the successful formulation of approximations that go beyond finite dimensional truncations in such a gravitational setting. The research reported in this paper addresses both of these challenges within the simplified setting of the conformally reduced quantum gravity model.

An appropriate technique to probe the non-perturbative regime of quantum gravity is provided by the functional RG expressed in terms of the effective average action (from here on simply ``effective action'') \cite{Wetterich:1992,Morris:1993}, a framework that has been applied in a wide variety of different contexts, \eg \cite{Berges:2000,Gies:2006wv,Pawlowski:2005xe,Morris:1998}. It is based on the idea of fully integrating out only those modes in the path integral that possess momenta larger than $k$, viewed as an infrared cutoff scale, whereas modes below this scale are suppressed via a momentum dependent mass term. By varying 
$k$, a non-perturbative RG flow is generated with the property that for $k \rightarrow 0$ the information contained in the full path integral is recovered.

Beginning with \cite{Reuter:1996}, there is a wealth of literature applying this idea to quantum gravity. For reviews and introductions see \cite{Reuter:2012,Percacci:2011fr,Niedermaier:2006wt,Nagy:2012ef,Litim:2011cp}. These works are aimed at investigating the hypothesis of asymptotic safety in quantum gravity \cite{Weinberg:1980}, which is the possibility that a continuum limit could be supported by an appropriate ultraviolet fixed point. However we emphasise that, whether or not this hypothesis is ultimately correct, it is important to make progress with the foundational, and computational, issues we are addressing here. In a separate companion paper \cite{DietzMorris:2015-2}, we investigate specifically the implications for asymptotic safety that follow from the flow equations we derive here.

Within the background field method,
an essential step in setting up the non-perturbative RG flow for gravity consists in splitting the metric field - in this framework this is the fundamental variable on which the quantum field theory of gravity is based - into a background metric $\bgmn$ and a fluctuation field $\thmn$: 
\be
\label{equ:metricsplit}
\tgmn = \bgmn + \thmn.
\ee
It is the fluctuation field $\thmn$ that is quantised, \ie it is being integrated over in the functional integral. The use of the background field method has the consequence that many conventional tools of quantum field theory become applicable for the derivation of the (functional RG) flow equation, \cf \cite{Reuter:1996}. In this way it is possible, for instance, to retain background gauge invariance for the effective average action which entails that it is a diffeomorphism invariant functional of its arguments. Furthermore, and crucially, the covariant derivative associated with the background field can now be used to define a momentum scale for fluctuation field modes, which is then compared to the cutoff scale $k$ to decide whether this mode is to be integrated out or suppressed.\footnote{We expand further on this below \eqref{equ:flowGamma}, and later comment on contributions from gauge fixing.}

At the same time it is clear that a background field split as in \eqref{equ:metricsplit} must not affect the physical observables themselves, in the sense that physical results have to be fully independent of this arbitrary choice of background metric $\bgmn$. Separate dependence on the background field may (and is bound to) appear at intermediate stages of the calculation but is required to disappear at the very end, \eg in the evaluation of Green's functions.

At the level of the effective action  in terms of which the flow equation is formulated, the background field appears as a separate argument alongside the full metric, $\Gamma_k[\gmn,\bgmn]$, where we have introduced the (total) classical metric $\gmn = \langle \tgmn \rangle$ and made the dependence on the RG scale $k$ explicit.  This setup and the ensuing flow equation for the effective action in principle 
does not rely on properties of a particular background field $\bgmn$ \cite{Reuter:1996}, implying that no background configuration assumes a distinguished role in this approach of quantising gravity and that effectively the quantisation proceeds on all backgrounds simultaneously (see \cite{Reuter:2008wj,Reuter:2008qx,Manrique:2010am,Manrique:2009uh}), 
at least up to choice of topology.\footnote{In practice simultaneous quantisation  is achieved through heat kernel (\ie short distance) expansions,  but exact evaluation of the functional trace in the flow equation requires specification of the  space-time topology  \cite{Demmel:2014fk,DietzMorris:2013-2}.} But making sure that the formalism does not rely on a particular background field is only the first step towards background independence of physical observables since it would generally still be the case that each background field configuration leads to \emph{different} observables due to the additional background field dependence in a  solution 
$\Gamma_k[g,\bar g]$ of the flow equation.
Instead, the requirement of background independence of physical observables is a further condition on the effective action that needs to be imposed separately and, as emphasised in ref. \cite{Bridle:2013sra}, in addition to the flow equation.

The majority of work on 
quantum gravity through this approach, has however been carried out in the so-called single field approximation. This approximation is characterised by essentially only retaining the $\bar g$-dependence contained in the classical gauge fixing and ghost terms and the identification $g= \bar g$ in $\Gamma_k[g,\bar g]$ at a certain stage of the calculation, \cf \cite{Becker:2014qya}. Background independence however can only be investigated in bi-field truncations, \ie truncations that retain dependence on both the full metric and the background metric. For studies going significantly beyond the single field approximation in different ways see \cite{Becker:2014qya,Manrique:2010am,Manrique:2009uh,Manrique:2010mq,Codello:2013fpa,Christiansen:2014raa,Groh:2010ta,Eichhorn:2010tb,Dona:2013qba,Dona:2014pla} and also for implications centered around the anomalous dimension of Newton's constant see \cite{Becker:2014jua,Dona:2013qba,Dona:2014pla,Falls:2015qga}.
Notably, it has recently been shown how background independence as a condition in this strict sense can be recovered in a double Einstein-Hilbert truncation  such that it becomes satisfied at the $k\to0$ end-point of the flow \cite{Becker:2014qya}.

One of the main aims of this work is to investigate background independence in the context of conformally reduced quantum gravity and to show 
that there exists an alternative, surely deeper, description in which background independence can actually be implemented exactly along the entire RG flow, through transformations which are deduced from combining the flow with the appropriate broken Ward Identities. 
Furthermore this will be implemented in an approximation that goes beyond polynomial truncations, 
keeping all powers of the fields, affording considerably greater insight into the structural issues.

We arrive at conformally reduced quantum gravity
by writing
\be
\label{equ:confpar}
\tgmn = f(\chi + \tilde \vp)\, \hat g_{\mu\nu} \qquad \text{and} \qquad \bgmn = f(\chi)\, \hat g_{\mu\nu},
\ee
where we consider only metrics that are conformally equivalent to the fixed reference metric $\hat g_{\mu\nu}$. These definitions imply  
that we can view $\chi(x)$ as the background conformal factor and $\tilde \vp(x)$ as the quantum conformal fluctuation field, making up the total quantum conformal factor $\tilde \phi(x) = \chi(x) + \tilde \vp(x)$. A path integral quantisation of $\tilde \vp$ will then lead to the classical fluctuation field $\vp = \langle \tilde \vp \rangle$ and total classical field $\phi = \langle \tilde \phi \rangle = \chi + \vp$. The quantum fluctuation field of the metric $\thmn = \tgmn-\bgmn$ is then given by $\thmn = \left[f(\tilde\phi)-f(\chi)\right]\hat g_{\mu\nu}$. Note that the difference on the right hand side may no longer be positive, corresponding to the fact that $\thmn$ need not be a metric field. 

Significantly, as we discuss later, we will find that we can work with a general parametrisation of the conformal factor as encoded in the function $f$.
In order to represent a valid parametrisation, 
the only requirement we need impose on $f(\phi)$ is that it be bijective, 
which means in this case that $f$ be  monotonic in $\phi$, and for a specified range of $\phi$, map to all positive real numbers.\footnote{As we will see, eventually the equations become independent of the choice of $f$, and thus we can formally also incorporate the singular case of vanishing metric ($f=0$) without any apparent difficulty.} 
Previously used choices include 
$f(\phi) = \exp(2\phi)$ \eg \cite{Machado:2009ph} where these conditions indeed hold, or $f(\phi)=\phi^2$,  \eg \cite{Manrique:2009uh,Bonanno:2012dg} where  to avoid double counting, arguably $\tilde\phi$ ought to be restricted to be positive (without loss of generality). 

Keeping only the conformal factor of the metric of course destroys diffeomorphism invariance. Nevertheless we are able to preserve a remnant of diffeomorphism invariance which will play an important r\^ole, in particular it will continue to strongly constrain the background field dependence. This remnant is multiplicative rescaling, and via the background field method the invariance is preserved by imposing
\be
\label{remnant}
x^\mu \mapsto x^\mu/\lambda\,,\qquad f(\chi)\mapsto\lambda^2 f(\chi)\,,
\ee
where the second equation follows from the expected transformation property of $\bgmn$, keeping the fixed reference metric fixed. By implementing the broken split Ward identity, we will in effect ensure that this remnant diffeomorphism invariance is inherited by the total field $\phi$ in the limit $k\to0$.

However in the ensuing we need to regard $\vp$ (and $\tilde\vp$) as a scalar field under \eqref{remnant}  to allow the cutoff term \eqref{equ:cutoff-action} to keep its simple bilinear form and at the same time have the correct transformation properties.
We can only achieve this in the exponential parametrisation. Then the requirement that  $f(\phi) = \exp(2\phi)$ also transforms as in \eqref{remnant}, can be solved with the background field covariant choice ($\chi\mapsto \chi+\ln\lambda$, $\phi\mapsto \phi+\ln\lambda$, $\vp\mapsto\vp$), which indeed is preserved by a background covariant cutoff. We can also realise this as a quantum covariance ($\chi\mapsto \chi$, ${\tilde\phi}\mapsto{\tilde \phi}+\ln\lambda$, ${\tilde\vp}\mapsto{\tilde\vp}+\ln\lambda$) which however is then broken by the cutoff, in close analogy to the properties of the background field formalism in the full theory.

For a general parametrisation $f$,  we can implement \eqref{remnant} by requiring all the conformal factor fields ($\phi,\vp,\chi$) to stay fixed, regarding the transformation \eqref{remnant} as acting only on $f$ and not its argument.  Although it is no longer a symmetry of any particular action (since the form of $f$ now changes), it is a symmetry of theory space which we can insist is preserved by our RG flow equations. It thus imposes powerful constraints which moreover agree with what we would expect to inherit from diffeomorphism invariance in the full theory.

By specialising to a background metric ${\bar g}_{\mu\nu}$ that is slowly varying (and thus also $\hat g_{\mu\nu}$), so that space-time derivatives of this can be neglected, we
effectively terminate at the level of the Local Potential Approximation (LPA) for the background conformal factor $\chi$. For the classical fluctuating conformal factor $\vp$ we will however fully implement  $\mathcal{O}(\partial^2)$ in the derivative expansion approximation.  We will make no other approximations.
In the conformally reduced model, the effective action will thus take its most general form at this level of truncation:
\be
\label{eff-ac-intro}
\Gamma_k[\vp,\chi] = \int d^dx 
\sqrt{\bar g}\left(-\frac{1}{2}K(\vp,\chi)\,{\bar g}^{\mu\nu} \partial_\mu \vp \partial_\nu\vp
		      +V(\vp,\chi)\right)\,,
\ee
where $V$ and $K$ are composite scalar fields to be determined.
When compared to the usual Einstein-Hilbert action, $K$ has absorbed the factor of $1/8\pi G$ and $V$ absorbed $\Lambda/16\pi G$ ($G$ being Newton's constant and $\Lambda$ the cosmological constant). The wrong-sign in the kinetic term is inherited from the infamous conformal factor problem 
\cite{Gibbons:1978ac}. The appearances of the background metric, and thus  background conformal factor $f(\chi)$, are as would be dictated by background diffeomorphism invariance (background curvature terms being neglected by the choice of slow $\bgmn$). Although diffeomorphism invariance is broken, at the LPA level within which we are working, remarkably both here and later the remnant symmetry \eqref{remnant} is enough to play the same r\^ole in that it constrains completely all appearances of $f(\chi)$. (This is particularly self evident when we specialise the ansatz above to \eqref{equ:ansatzGamma}, following the choice ${\hat g}_{\mu\nu}=\delta_{\mu\nu}$.)

The effective action (where we have omitted a parametric dependence on the fixed reference metric $\hat g$)  satisfies the flow equation (\cf sec.\ref{sec:flowequ-derivation})
\be
\label{equ:flowGamma}
\partial_t \Gamma_k[\vp,\chi] = \frac{1}{2}\mathrm{Tr}\left[\frac{1}{\sqrt{\bar g}\sqrt{\bar g}}\frac{\delta^2\Gamma_k}
				  {\delta \vp \delta \vp}+ R_k[\chi]\right]^{-1} \partial_t R_k[\chi]\,.
\ee
(Since local diffeomorphism invariance is broken by conformal reduction, gauge fixing and ghosts are not required.) 
 Here we have introduced the RG time 
\be
\label{time}
t=\ln(k/\mu)\,,
\ee 
with $\mu$ being a fixed reference scale, which can be thought of as being the usual arbitrary finite physical mass-scale. $R_k$ is the cutoff operator responsible for suppressing momentum modes below the infrared cutoff scale $k$, \cf \cite{Wetterich:1992,Morris:1993}.

The crucial observation is that in the context of the background field method in quantum gravity the cutoff operator itself depends on the background field $\chi$. The reason for this is that the cutoff operator is a function of the covariant Laplacian of the background metric $R_k\left(-\bar \nabla^2\right)$, as it is with respect to the spectrum of $-\bar\nabla^2$ that modes are integrated out or suppressed in the path integral, \cf \cite{Reuter:2008wj,Reuter:2009kq}. Again we emphasise that the remnant diffeomorphism invariance \eqref{remnant} enforces precisely this $\chi$ dependence at the LPA level (given that this cutoff term is to be bilinear in the quantum field $\tilde\vp$). In turn it is this extra background field dependence of the cutoff that forces the effective action $\Gamma_k[\vp,\chi]$ to have separate background field dependence. 

It is possible to keep track of the background field dependence through an identity similar to the flow equation, \cite{Pawlowski:2005xe,Litim:2002hj,Bridle:2013sra,Reuter:1997gx,Litim:1998nf,Litim:2002ce,Manrique:2009uh,Manrique:2010mq,Manrique:2010am} 
which in the present context assumes the following form,
\be
\label{equ:sWiGamma}
\frac{1}{\sqrt{\bar g}}\left(\frac{\delta\Gamma_k}{\delta \chi}-\frac{\delta \Gamma_k}{\delta \vp}\right)
      =\frac{1}{2}\mathrm{Tr}\left[\frac{1}{\sqrt{\bar g}\sqrt{\bar g}}\frac{\delta^2\Gamma_k}
				  {\delta \vp \delta \vp}+ R_k[\chi]\right]^{-1} 
				  \left\{\frac{1}{\sqrt{\bar g}}\frac{\delta R_k[\chi] }{\delta \chi}+Y_f R_k[\chi]\right\}.
\ee
For reasons detailed in sec. \ref{sec:sWI-derivation}, we will refer to this equation as the broken, or modified, split Ward Identity (msWI). Exact background independence would be realised if the right hand side of the msWI was zero, implying that the effective action is only a functional of the total field $\phi = \chi + \vp$. The presence of the cutoff operator however causes the right hand side to be non-vanishing in general. It is only in the limit $k\rightarrow0$ (holding physical, \ie unscaled, momenta and fields fixed) that the cutoff operator drops out and background independence can be restored exactly. We note therefore that imposing the msWI in addition to the flow equation \eqref{equ:flowGamma} automatically ensures exact background independence in the limit $k\rightarrow0$. The multiplicative function $Y_f$ in \eqref{equ:sWiGamma} is connected to the parametrisation of the conformal factor in \eqref{equ:confpar} and will be determined explicitly in sec. \ref{sec:sWI-derivation}.

Even though at any finite $k>0$ exact background independence will inevitably be lost due to the cutoff, the msWI represents the natural continuation of full background independence and it is necessary to impose this to restrict the separate background field dependence of the effective action. Recalling an argument in ref. \cite{Bridle:2013sra},
this can easily be seen by modifying a solution $\Gamma_k[\vp,\chi]$ of the flow equation \eqref{equ:flowGamma} by an arbitrary \emph{scale independent} functional of the background field $\Gamma_k[\vp,\chi]+\mathcal{F}[\chi]$. The result would clearly still be a solution of the flow equation,  a small reflection of the additional freedom available through the separate background field dependence of the effective action. Solutions of the flow equation \eqref{equ:flowGamma} therefore do not automatically solve the msWI \eqref{equ:sWiGamma}, but the latter instead controls the enlargement of theory space introduced by hand through the background field split \eqref{equ:metricsplit} or \eqref{equ:confpar}, which must of course not affect the physical results that we eventually aim to obtain from $\Gamma_k$. 

It is worth emphasising that the msWI constraint \eqref{equ:sWiGamma} is not an optional extra. Since it follows directly from the form of the partition function and the infrared cutoff operator (\cf \eqref{equ:pathint-conffact} and sec. \ref{sec:flowequ-derivation}), it has to be satisfied at all points on the flow. For the truncation considered in this work we will show how the msWI can be imposed alongside the flow equation for any value of $k$. In doing so it will become clear how the msWI takes appropriate care of the separate background field dependence of $\Gamma_k[\vp,\chi]$ all along the RG flow and in particular in the limit $k\rightarrow 0$ implements exact background independence.

When all degrees of freedom of the metric are considered in full gravity, the msWI \eqref{equ:sWiGamma} receives additional terms on its right hand side. Beyond the analogue of \eqref{equ:sWiGamma} for the metric fluctuation field $\hmn$ it contains terms originating from background dependence induced through the choice of background-covariant gauge fixing and from cutoff terms in the  ghost action itself. 
Nevertheless, the above discussion carries over to full gravity. In the limit $k\rightarrow 0$ all cutoff terms are required to vanish and exact background independence can in principle be restored by noting that the gauge fixing and ghost terms on the right hand side of the msWI are expected to drop out upon going ``on-shell'' (assuming such an appropriate property may be defined). The recent studies \cite{Becker:2014qya,Codello:2013fpa,Christiansen:2014raa} indicate that disentangling background field effects from the true physical content of the effective action leads to improved results compared to the use of the above mentioned single field approximation (see also \cite{Demmel:2014hla} for improvements within the single field approximation). A specific example where serious problems are encountered in the single field approximation is provided by the so-called $f(R)$ truncation, \cf \cite{Benedetti:2012,Codello:2008,Machado:2007}. This truncation has been shown to break down in the sense that the physical spaces of eigenoperators are empty due to a reparametrisation redundancy in the flow equation \cite{DietzMorris:2013-2}. The results of the study \cite{Bridle:2013sra} in particular point towards the possibility that these problems could be resolved in a more comprehensive bi-metric approach that makes use of the msWI \eqref{equ:sWiGamma}. 

In ref. \cite{Bridle:2013sra} these issues were investigated in a toy model, namely scalar field theory at the level of the LPA. Incorporating by hand some background dependence in the effective cutoff, it was demonstrated how this could then be removed by imposing also the msWI, leading to a different formulation with explicit background independence along the entire RG flow. Those results inspired the research reported here. As we have just reviewed, providing we impose the msWI, general arguments should ensure that in the gravitational case we recover background independence in the limit $k\to0$. However, given that diffeomorphism invariance does not allow the background dependence to be removed from the cutoff, there is no reason {\it a priori} to expect that a background independent description of the \emph{entire} RG flow should be possible. From this viewpoint our discovery that such a thing does exist, at least within the model approximation studied here, seems to us both dramatic and already to hint at some deeper understanding of the meaning of the RG in quantum gravity. 

As an additional result that becomes possible through the use of the msWI \eqref{equ:sWiGamma} we will show in sec. \ref{sec:independence} that the specific form in which the conformal factor is parametrised in \eqref{equ:confpar} as given by the function $f$ has no influence on this deeper description. Since the solutions of the background independent RG equations are mapped in one-to-one fashion into solutions of the original flow equations, the incorporation of the msWI thus also ensures that the quantisation of the conformal factor  is independent of 
its parametrisation,\footnote{at least in the truncation
considered here, and up to only topological considerations of the range of $\phi$} another dramatic result which, whilst it could have been expected or at least desired from a physical point of view, has not been achieved before.

As discussed at the very beginning, the most profound property which we are forced to confront is the inherent ambiguity in the meaning of RG scale, closely bound up in this framework with background dependence.  
Since $k$ is defined with respect to modes of the covariant Laplacian $-{\bar\nabla}^2$, or more generally tied to some background measure of inverse length, it is inherently background field dependent. This notion of RG scale must therefore be replaced by some other notion if the end result is to be a system of flow equations that is completely background independent. We will see in sec. \ref{sec:independence} that such a notion is uniquely and self-consistently determined from the structure of the equations, and is encapsulated in a new quantity\footnote{$\alpha$ is a function of scaling dimensions, \eqref{eq:alpha}, and $d_f$ is the scaling dimension of $f$} 
\be
\label{khat}
\kh = k^\frac{\alpha}{\alpha+d_f}f(\chi)^\frac{1}{\alpha+d_f}\,.
\ee

As we will see in sec. \ref{sec:fpmap}, fixed points with respect to $\hat k$ do not have to bear any relation to fixed points with respect to $k$. This is to be expected since $\hat k$-fixed points are explicitly background independent, whereas as discussed below \eqref{equ:sWiGamma}, $k$-fixed points can only be background independent in general using fixed physical units and taking the limit $k\to0$. Worse than that we will see that in general $k$-fixed points are actually forbidden to exist by this requirement. 
However we will also see that $k$-fixed points may exist in general for power law $f(\phi)\propto \phi^\gamma$,  and moreover for judicious choices of the form of $f$ and its scaling dimension $d_f$, the two notions of fixed points can coincide.

The results we have been describing were reached by being guided by the mathematical structure of the equations, and  this is how we will present them  in secs. \ref{sec:flowequ-derivation}--\ref{sec:combine}. It is important that progress can be made in this way since guidance from the mathematical structure will become even more important in fully fledged quantum gravity. But on the other hand the results then seem quite obscure from the physical point of view. 

In sec. \ref{sec:dimensions}, we aim to rectify this as much as possible by pointing out a number of subtle issues that underly the equations. In particular it is important to note that the background metric ${\bar g}_{\mu\nu}$ receives its definition in the ultraviolet, as part of the bare action and functional integral. This includes for example its r\^ole in the source term and insertion of the infrared cutoff \cf eqns. (\ref{equ:pathint-conffact}-\ref{equ:source}). Through the derivation (see secs. \ref{sec:flowequ-derivation} and \ref{sec:sWI-derivation}), it is this {bare} ${\bar g}_{\mu\nu}$ that then appears explicitly in the flow equation \eqref{equ:flowGamma} and msWI \eqref{equ:sWiGamma}. 
This matters because nothing prevents the conformal factor from picking up an anomalous scaling dimension $[\phi]=\eta/2\ne0$, as we will see explicitly in secs. \ref{sec:independence} and \ref{sec:dimensions}. But then $f(\phi)$ in general must contain a bare dimensionful parameter.  This has two profound consequences. Firstly $f$ must then run with scale, which already suggests that the two notions of fixed point cannot coincide, and secondly it means that the quantities in \eqref{eff-ac-intro} need to be renormalised. Fortunately, as we will see in sec. \ref{sec:dimensions}, the existence of background independent variables allows us to pass from  bare to renormalised quantities in such a way that the effective action takes the same form as \eqref{eff-ac-intro} but with all quantities replaced by their renormalised counter-parts.

At first sight we have no choice 
over the dimensional assignment of $[f]=d_f$
since if we regard the reference metric $\hat g_{\mu\nu}$ as on the same footing as the quantum metric $\tgmn$ and background metric $\bgmn$, then we must set $d_f=0$ as is clear from \eqref{equ:confpar}. However, multiple choices of dimensional assignments necessarily co-exist in the theory. For example, the remnant diffeomorphism invariance \eqref{remnant} can itself be regarded as a particular choice of dimensional assignments $[\ ]=[\ ]_D$ where we set $[x]_D=-1$ and $[f]_D=2$, and all other dimensions are set to zero. Linear combinations of these dimensional assignments and the usual assignments are then also allowed, corresponding to a mixing between different scaling symmetries. As fully explored in sec. \ref{sec:dimensions}, we will see in this way why a general dimension $d_f$ can be incorporated, and why specific choices then allow the fixed points with respect to $k$ and $\hat k$ to coincide.

As already mentioned, we defer to a companion paper \cite{DietzMorris:2015-2} a detailed analysis of the properties of fixed points of the resulting background independent flow equations.

The structure of the rest of the paper is as follows. In sec. \ref{sec:BeyondLPA}, generalising \cite{Morris:1994ie}, we formulate in an elegant way the technical steps necessary to compute the functional trace in both the flow equation and msWI, algorithmically, order by order in a derivative expansion, whilst carefully negotiating some subtleties of the slow background field limit. In sec. \ref{sec:comp-sft} we compare the background-independent flow equations to those from scalar field theory at $\mathcal{O}(\partial^2)$ level, highlighting both the close similarities at this stage but also the crucial differences. In sec. \ref{sec:earlier} we demonstrate why differences in our approach make no direct comparison possible with earlier work. 
Finally, in sec. \ref{sec:conclusions}, we summarise and draw our conclusions.

\section{The flow equation for the conformal factor}
\label{sec:flowequ-derivation}
Truncating theory space to contain functionals of the conformal factor only by writing \eqref{equ:confpar} leaves us to consider effective actions that depend on the two single-component fields $\vp$ and $\chi$. This simplifies the flow equation for $\Gamma_k$ as we no longer have to take into account any gauge fixing or ghost terms, since these degrees of freedom are discarded by our insistence on a fixed reference metric ${\hat g}_{\mu\nu}$ in \eqref{equ:confpar}. Nevertheless, the flow equation does not simply reduce to that of scalar field theory in the presence of a background field but still contains important features indicating its origin from a flow equation on a truncated theory space of quantum gravity. In order to spell out these differences we go through the derivation of \eqref{equ:flowGamma} from the path integral in some detail.

The starting point is the Euclidean functional integral over the fluctuation part of the conformal factor,
\be
\label{equ:pathint-conffact}
\exp(W_k) = \int  \mathcal{D}\tilde\vp \, \exp\left(-S[\chi + \tilde \vp] - S_k[\tilde \vp,\bar g] + S_\mathrm{src}[\tilde \vp,\bar g]\right),
\ee
where the quantum analogue of the classical $\vp$ is denoted $\tilde \vp$. In order to make sense of this construct we need to provide an overall ultraviolet regularisation at some high scale $k_0$. We will later need to acknowledge dependence on this scale $k_0$, but we do not need to specify the ultraviolet regularisation explicitly.\footnote{The simplest and consistent choice however is to the use the Legendre transform relation in ref. \cite{Morris:1993} to define the bare action as the resulting Wilsonian effective action at the scale $k_0$ \cite{MorrisSlade}.}
In this equation and in what follows we will continue to express quantities in terms of the background metric $\bar g$ to highlight the fact that the results are those that would be expected more generally in a diffeomorphism invariant theory, but underneath these appearances lies the identification as in the second equation of \eqref{equ:confpar}. We will make all this fully explicit later. The implied appearances of $f(\chi)$ are in fact completely dictated by the remnant background diffeomorphism invariance \eqref{remnant} as we will see, and as already emphasised in the Introduction. 

Under the conformal truncation \eqref{equ:confpar}, the bare action for gravity turns into the bare action for the total conformal factor, 
\begin{equation} \label{equ:confBareAction}
S[\bar g + \tilde h] \rightarrow S[\chi +\tilde \vp].
\end{equation}
The cutoff action $S_k$ has to suppress the momentum modes of the dynamical field $\tilde \vp$ and takes the form
\be
\label{equ:cutoff-action}
S_k[\tilde \vp,\bar g] = \frac{1}{2}\int d^dx \sqrt{\bar g(x)}\, \tilde \vp(x) R_k[\bar g] \tilde \vp(x),
\ee
where we are operating in $d$ dimensions and have explicitly kept the Riemannian measure of the background field in the same way as it appears in the cutoff action of the full gravitational functional integral $S_k[\thmn,\bgmn]$, \cf \cite{Reuter:1996}. This is in agreement with the background field method \cite{Abbott:1980hw,Abbott:1981ke}: the gauge degrees of freedom should be absorbed by $\bgmn$, leaving $\tilde\vp$ to transform covariantly, in this case as a scalar field. $R_k[{\bar g}]$ is to be regarded as a scalar operator (\ie mapping scalar fields to scalar fields). Then we note that as advertised, the remnant diffeomorphism invariance \eqref{remnant} is in fact sufficient on its own to fix the $f(\chi)$ dependence of the measure in \eqref{equ:cutoff-action}. We have also made explicit the dependence of the cutoff operator on the background metric. Again bearing the gravitational setting in mind, and in agreement with \eqref{remnant}, the source term needs to be of form
\be
\label{equ:source}
S_\mathrm{src}[\tilde \vp,\bar g] = \int d^dx\sqrt{\bar g(x)}\, \tilde \vp(x)J(x).
\ee
Assuming that $\bgmn$ is $k$ independent, taking a $t$-derivative of \eqref{equ:pathint-conffact} leads to
\be
\label{equ:dtWk}
\partial_t W_k = -\frac{1}{2}\int d^dx\sqrt{\bar g(x)}\, d^dy\sqrt{\bar g(y)}\, \partial_t R_k[\bar g](x,y) 
		  \langle \tilde \vp(x) \tilde\vp(y)\rangle,
\ee
where $t$ is defined by \eqref{time} and we have rewritten the cutoff operator using its kernel according to
\be
\label{equ:kernel-notation}
R_k(x,y) = R_{k,x}\, \frac{\delta(x-y)}{\sqrt{\bar g(y)}},
\ee
with the subscript in $R_{k,x}$ indicating that this differential operator acts on $x$-dependence. We then proceed in the usual way via a Legendre transformation with respect to the source,
\be
\label{equ:LegendreTrans}
\tilde \Gamma_k[\vp,\bar g] = \int d^dx\sqrt{\bar g(x)} \, J(x)\vp(x) - W_k[J,\bar g],\qquad \text{with} \qquad
	\vp =\frac{1}{\sqrt{\bar g}}\frac{\delta W_k}{\delta J}
\ee
so that using the well-known identity
\be
\label{equ:rel-expval}
\langle\tilde\vp(x) \tilde\vp(y)\rangle = 
\left(\frac{1}{\sqrt{\bar g(x)}\sqrt{\bar g(y)}}\frac{\delta^2\tilde \Gamma_k}{\delta \vp(x)\delta\vp(y)}\right)^{-1} + \vp(x)\vp(y)
\ee
together with the redefinition $\Gamma_k[\vp,\bar g] = \tilde \Gamma_k[\vp,\bar g] - S_k[\vp,\bar g]$ turns \eqref{equ:dtWk} into
\be
\label{equ:flowGamma-detail}
\partial_t\Gamma_k[\vp,\bar g] = \frac{1}{2}\int_x \int_y
				  \left[\Gamma^{(2)}(x,y) + R_k(x,y)\right]^{-1} \, \partial_t R_k(y,x).
\ee
Here, we have abbreviated $\int_x \equiv \int d^dx\sqrt{\bar g(x)}$ and written 
\begin{equation}
\Gamma^{(2)}(x,y)=\frac{1}{\sqrt{\bar g(x)}\sqrt{\bar g(y)}}\frac{\delta^2 \Gamma_k}{\delta \vp(x)\delta\vp(y)}
\end{equation}
for the Hessian of the effective action. We have further exploited the fact that the kernel $R_k(x,y)$ is symmetric. The flow equation \eqref{equ:flowGamma-detail} is the detailed form of \eqref{equ:flowGamma} and it contains the appropriate factors of the background field in places where they are to be expected in a truncation of a gravitational effective action, while the implied factors of $f(\chi)$ must appear in these places as imposed by the remnant background diffeomorphism symmetry \eqref{remnant}.

As remarked in the Introduction, in these equations and thus also in what follows, it is clear that $\bgmn$ is a bare quantity. This will become important when we acknowledge that through \eqref{equ:confpar} it becomes a composite operator. For now all we need is the fact that the relations \eqref{equ:confpar} are therefore $k$ independent, as was assumed in deriving \eqref{equ:dtWk}. We can safely ignore the fact that $f$ generically depends on the overall cutoff scale $k_0$, until sec. \ref{sec:dimensions} when we will show how the equations can be renormalised.

For completeness, let us stress that because of the replacement \eqref{equ:confBareAction} in the functional integral \eqref{equ:pathint-conffact} using \eqref{equ:confpar}, the effective action in \eqref{equ:flowGamma-detail} of course has a dependence on $\chi$ that we have not indicated notationally. Later on we will make the identification $\bgmn = f(\chi)\, \hat g_{\mu\nu}$ so that the second argument of the effective action $\Gamma_k[\vp,\bar g]$ turns into the background conformal factor $\chi$, making the dependence on  $\chi$ explicit.

Connected with the conformal truncation \eqref{equ:confBareAction}, there is another issue worth mentioning. Any truncation ansatz for the effective action $\Gamma_k[\vp,\bar g]$ in \eqref{equ:flowGamma-detail} that reflects its gravitational nature will contain the parametrisation function $f$ entering through the second equation in \eqref{equ:confpar}, as enforced by \eqref{remnant}, \cf \eqref{eff-ac-intro} or more explicitly \eqref{equ:ansatzGamma}. As mentioned in the Introduction we will show in sec. \ref{sec:independence} that this dependence on the way the conformal factor is parametrised can be completely eliminated through the use of the msWI of the following section. In this sense the quantisation of the conformal factor, at least in the truncation we will consider, is independent of the parametrisation $f$. But there is a more subtle way in which the parametrisation function enters the current setup. Starting from a bare action of the metric field $S\left[\tgmn\right]=S[\bgmn + \thmn]$ and applying the conformal truncation \eqref{equ:confpar} leads to a bare action $S_f[\chi + \tilde \vp]$ that depends on the parametrisation function $f$. In principle this $f$-dependence carries through to the effective action in the flow equation \eqref{equ:flowGamma-detail} and this could in general be an additional dependence on $f$ that is not explicitly represented in any ansatz one might propose for $\Gamma_k[\vp,\bar g]$. However we can see this is not so, via a universality argument. If we formally imagine the effective action to be expanded in terms of all allowed operators,
\begin{equation} \label{equ:GammaOpExp}
\Gamma_k[\vp,\bar g] = \sum_n g_n(k)\, \mathcal O_n[\vp,\bar g]
\end{equation}
with RG scale dependent couplings $g_n(k)$, there will be a map between the couplings of this expansion and the bare couplings contained in the bare action $S[\tgmn] = S[f(\chi+\tilde\vp)\hat g_{\mu\nu}]$ as given by the Legendre transform \eqref{equ:LegendreTrans}. Adopting a different choice of parametrisation function $f$ in \eqref{equ:confpar} will only lead to a correspondingly modified map between the bare couplings and the couplings in the effective action \eqref{equ:GammaOpExp}. In other words, if we can find a satisfactory solution of the flow equation \eqref{equ:flowGamma-detail} different parametrisations $f$ simply correspond to different ways of mapping the 
couplings of the effective action onto the 
couplings of the bare action in the functional integral approach. Given that we can derive any physical observable from the effective action, we may regard any $f$-dependence that affects only the bare action couplings in this way as physically irrelevant.

\section{The split Ward identity for the conformal factor}
\label{sec:sWI-derivation}
The background field split for the full metric \eqref{equ:metricsplit} carries over to the analogous background field split of the conformal factor in \eqref{equ:confpar} and it is a characteristic feature of gravity that the effective action possesses a separate dependence on the background field. This is the case for full gravity but also in the conformal truncation. As mentioned in the Introduction this comes especially from the fact that the coarse graining procedure as implemented by the cutoff operator $R_k$ is realised through the covariant background Laplacian $-\bar \nabla^2$, which in the conformal approximation \eqref{equ:confpar} results in a dependence of the cutoff on the background conformal factor $\chi$, as dictated by \eqref{remnant}. In order to keep track of this background field dependence and to ensure background independence in the limit $k \rightarrow 0$ we need the msWI \eqref{equ:sWiGamma}, the derivation of which is given here.

For the sake of clarity it will be useful in this section to make the identification 
\begin{equation}\label{equ:choiceBackg}
\bgmn=f(\chi)\,\delta_{\mu\nu}
\end{equation}
of \eqref{equ:confpar} with the fixed reference metric taken to be flat. This choice will be used in actual computations in the later sections of this work.

The functional integral for the conformal factor \eqref{equ:pathint-conffact} then becomes
\be
\label{equ:pathint-conffact-sWI}
\exp(W_k) = \int  \mathcal{D}\tilde\vp \, \exp(-S[\chi + \tilde \vp] - S_k[\tilde \vp,\chi] + S_\mathrm{src}[\tilde \vp,\chi]).
\ee
The msWI originates from the observation that the bare action in this partition function is invariant under split symmetry (called shift invariance in ref. \cite{Bridle:2013sra}) as expressed in
\be
\label{equ:split-symmetry}
\tilde \vp(x) \mapsto \tilde \vp(x) + \eps(x) \qquad \chi(x) \mapsto \chi(x) -\eps(x),
\ee
but the cutoff action breaks this invariance,\footnote{The source term also breaks split invariance but does not contribute to the separate background field dependence in $\Gamma_k[\vp,\chi]$, as the ensuing calculation shows.}  \cf \cite{Becker:2014qya,Reuter:2008qx,Bridle:2013sra}. It is the violation of split symmetry that signals the loss of background independence, both at the level of the functional integral and at the level of the effective action.

Applying the shifts \eqref{equ:split-symmetry} with an infinitesimal $\eps(x)$ to the functional integral \eqref{equ:pathint-conffact-sWI} with $\sqrt{\bar g}=f(\chi)^{d/2}$ in the cutoff action \eqref{equ:cutoff-action} and the source term \eqref{equ:source} leads to
\be
\label{equ:varW-sWI}
-\int_w f^{-\frac{d}{2}}\frac{\delta W_k}{\delta \chi} \,\eps = \int_w \left[
		J -\frac{d}{2}\dclnf \langle \tilde\vp\rangle J 
		-R_k \langle \tilde\vp\rangle +\frac{d}{4}\dclnf\langle \tilde \vp \,R_k \tilde \vp
		\rangle + \frac{1}{2}\langle \tilde \vp \,\partial_\chi R_k \,\tilde \vp\rangle\right]\eps.
\ee
The Legendre transformation \eqref{equ:LegendreTrans} with the shifts \eqref{equ:split-symmetry} gives
\be
\int_w f^{-\frac{d}{2}}\frac{\delta W_k}{\delta \chi} \, \eps = 
		\int_w \left[\frac{d}{2}\left(\dclnf J\vp\right) - \right.
		\left. f^{-\frac{d}{2}}\frac{\delta \tilde \Gamma_k}{\delta \chi} \right] \eps,
\ee
which we use in \eqref{equ:varW-sWI}. Performing similar steps as in the previous section, exploiting \eqref{equ:rel-expval} and redefining $\Gamma_k[\vp,\chi] = \tilde \Gamma_k[\vp,\chi] - S_k[\vp,\chi]$ then results in
\begin{equation}
\label{equ:sWI-detail-eps}
\int_w f^{-\frac{d}{2}}\left(\frac{\delta \Gamma_k}{\delta \chi}-\frac{\delta \Gamma_k}{\delta \vp}\right)\eps =
 \frac{1}{2}\int_x \int_y
 \Delta(x,y)\left[\frac{d}{2}\,\partial_\chi\! \ln\! f(y) \,R_k(y,x)+\partial_\chi R_k(y,x)\right]
 \eps(y).
\end{equation}
Here we have again used the kernel notation \eqref{equ:kernel-notation}, the integral notation introduced below \eqref{equ:flowGamma-detail} and abbreviated
\be
\label{equ:Delta}
\Delta(x,y) = \langle \tilde \vp(x) \tilde \vp(y)\rangle - \vp(x)\vp(y)
	    = \left[f(x)^{-\frac{d}{2}}f(y)^{-\frac{d}{2}}\frac{\delta^2 \Gamma_k}{\delta\vp(x)\delta\vp(y)}+R_k(x,y)\right]^{-1}.
\ee
The parametrisation function $f$ inherits its coordinate dependence from the background field $\chi$ and it is therefore useful use the shorthand $f(x) \equiv f(\chi(x))$, where appropriate. The msWI is obtained by dropping the arbitrary choice of $\eps(w)$,
\begin{equation}
\label{equ:sWI-detail}
f(w)^{-\frac{d}{2}}\left(\frac{\delta \Gamma_k}{\delta \chi(w)}-\frac{\delta \Gamma_k}{\delta \vp(w)}\right) =
 \frac{1}{2}\int_x \int_y
 \Delta(x,y)\left[\frac{d}{2}\,\partial_\chi\! \ln\! f(y) \,R_k(y,x)+\partial_\chi R_k(y,x)\right]
 \frac{\delta(y-w)}{f(w)^{\frac{d}{2}}}.
\end{equation}
We see that \eqref{equ:sWI-detail} is the written out form of \eqref{equ:sWiGamma}, where 
\be
Y_f(x,w) = \frac{d}{2}\dclnf\frac{\delta(x-w)}{f(w)^{\frac{d}{2}}}
\ee
is an additional contribution, beyond the background field dependence of the cutoff operator itself, that originates from the background field contained in the measure of the cutoff action \eqref{equ:cutoff-action}. We will see in sec. \ref{sec:independence} that both contributions are crucial for the implementation of the msWI \eqref{equ:sWI-detail} on the entire RG flow.

\section{Derivative expansion}
\label{sec:BeyondLPA}

In this section we develop all the steps necessary to perform a derivative expansion of the effective action, flow equation \eqref{equ:flowGamma-detail} and msWI \eqref{equ:sWI-detail}, specialised to the slow background field case.

\subsection{Derivative expansion of the effective action}
\label{sec:eff-ac}

In order to truncate the effective action of the conformal factor $\Gamma_k[\vp,\chi]$ we make use of a derivative expansion. This is an expansion scheme that has proved successful in applications of the functional RG to other quantum field theories such as scalar field theory, \eg \cite{Morris:1994ie}. As already mentioned in the Introduction we simplify matters by specialising to a slow background field such that $\partial_\mu\chi$ is neglected. We now also  specialise ${\hat g}_{\mu\nu}=\delta_{\mu\nu}$ as in \eqref{equ:choiceBackg}. Then our already advertised ansatz \eqref{eff-ac-intro} takes the form
\be
\label{equ:ansatzGamma}
\Gamma_k[\vp,\chi] = \int d^dx f(\chi)^\frac{d}{2}\left(-\frac{1}{2}{K(\vp,\chi)\over f(\chi)}\left(\partial_\mu \vp\right)^2
		      +V(\vp,\chi)\right)
\ee
in which we keep a general scalar potential $V(\vp,\chi)$ at zeroth order of the derivative expansion and a general scalar function $K(\vp,\chi)$ at $\mathcal{O}\big(\partial^2\big)$ for the fluctuation field $\vp$. It is understood that both of them depend on the RG time $t$. 

Since this effective action represents an approximation in the conformal sector of quantum gravity we want to see the appropriate measure factor $f^{d/2}$ as above. Similarly in order to keep as close as possible to the formulation in full quantum gravity, we want the kinetic term in \eqref{equ:ansatzGamma} to appear as above, since the $1/f$ factor is what remains from the invariant construction using ${\bar g}^{\mu\nu}$. 
Consistent with these expectations, these dependencies are in fact enforced by the remnant background diffeomorphism invariance \eqref{remnant}.

A further conspicuous feature of \eqref{equ:ansatzGamma} is the minus sign in front of the kinetic term of $\vp$, reflecting the negative sign of the kinetic term of the conformal factor that is obtained for the Einstein-Hilbert action and is otherwise known as the the conformal factor problem. Note that as far as the effective action is concerned, this sign can be accommodated in a natural way as already discussed in \cite{Reuter:1996}, although see also ref. \cite{Bonanno:2012dg} and our comments in sec. \ref{sec:comp-sft}. In the context of conformally reduced quantum gravity it has also been discussed in \cite{Manrique:2009uh,Reuter:2008qx,Reuter:2008wj}. We will come back to this in sec. \ref{sec:earlier}.

Of course a full treatment of the effective action at $\mathcal{O}\big(\partial^2\big)$ of the derivative expansion would also allow $\chi(x)$ to vary with $x$ such that 
 an analogous kinetic term for the background field can no longer be neglected and indeed one would then also have to include a mixed term $\sim \partial_\mu\vp\partial_\mu\chi$. We comment further on this in the Conclusions.

\subsection{Strategy for the evaluation of the traces}
\label{sec:strategy-traces}
An essential step in deriving the flow equation and the msWI for the ansatz \eqref{equ:ansatzGamma} is the evaluation of the traces in \eqref{equ:flowGamma-detail} and \eqref{equ:sWI-detail}. In the following we will derive a general framework that can be used to perform this task in the present context of conformally reduced gravity and for a derivative expansion of the effective action.

We are interested in the expression
\be
\label{equ:gen-trace}
T_\mathcal{K} = \int d^dxd^dy \sqrt{\bar g(x)}\sqrt{\bar g(y)}\, \Delta(x,y) \mathcal{K}(y,x)
\ee
for a generic kernel $\mathcal{K}$ that can be chosen correspondingly for the flow equation and the msWI, and $\Delta(x,y)$ has been introduced in \eqref{equ:Delta}.

For any function $u(x)$, the kernel $\mathcal{K}(x,y)$ satisfies the relation
\be
\int d^dy \sqrt{\bar g(y)}\,\mathcal{K}(x,y)u(y) = \mathcal{K}_x u(x),
\ee
where $\mathcal{K}_x$ is the associated differential operator acting on $x$-dependence. This relation between the kernel and its differential operator is satisfied if they obey \eqref{equ:kernel-notation} with $\mathcal{K}$ in place of $R_k$. Using this in \eqref{equ:gen-trace}, we find
\begin{equation}
\label{temp1}
T_\mathcal{K} = \int d^dx\, d^dy \, \left[\Delta_x\delta(x-y)\right]\left[\mathcal{K}_y \delta(y-x)\right]	
	      = \int \left. d^dx \,\Delta_x \, \mathcal{K}_x \,\delta(x-x')\right|_{x'=x},
\end{equation}
where in the last step the differential operators act on the delta function before setting $x'=x$. For the flow equation \eqref{equ:flowGamma-detail} we simply have $\mathcal{K}_x =  \partial_t R_{k,x}$ while for the msWI \eqref{equ:sWI-detail} the operator $\mathcal{K}_x$ takes the form
\be
\label{equ:KsWI}
\mathcal{K}_x = \frac{\delta(x-w)}{f(w)^\frac{d}{2}}\left\{\frac{d}{2}\dclnf R_{k,x} 
+\partial_\chi R_{k,x}\right\}.
\ee
The differential operator $R_{k,x}$ acts to its right, leaving the delta function untouched. It is a function of the covariant background field Laplacian,
\be
R_{k,x} = R_k\left(-\bar \nabla^2_x\right) = R_k\left(-f(\chi)^{-1}\partial^2_x \right)
\ee
where we have made use of the conformal reduction \eqref{equ:confpar} and neglected any derivatives of $\chi$. Using this, and by expressing $\delta(x-x')$ in momentum space, \eqref{temp1} can be transformed as follows,
\begin{align}
T_\mathcal{K} &= \int  d^dx \frac{d^dp}{(2\pi)^d} 
			\left. e^{-i p\cdot x'}\Delta_x \,
			 \mathcal{K}\left(-f^{-1}\partial_x^2\right)e^{ip\cdot x} \right|_{x=x'}	\\
			 &= \int  d^dx \frac{d^dp}{(2\pi)^d}
			 	\left\{e^{-i p\cdot x}\Delta_x e^{i p\cdot x}\right\} \mathcal{K}\left(f^{-1} p^2\right).
\end{align}
In the last step we have again neglected all $\partial_\mu \chi$-terms. Since we make use of this specialisation to slow background fields, the operator $\mathcal{K}$ in this calculation effectively has no separate dependence on $w$ or $x$ through the background field $\chi$. In sec. \ref{sec:evalsWI}, a careful treatment of this slow background field assumption demonstrates that this is also effectively true for the delta function appearing in \eqref{equ:KsWI}. Abbreviating 
\be
Q(p,x)=e^{-i p\cdot x}\Delta_x e^{i p\cdot x}
\ee
we can finally write
\be
\label{equ:traces}
T_\mathcal{K} = \int  d^dx \frac{d^dp}{(2\pi)^d} \,
			 	Q(p,x) \, \mathcal{K}\left(f^{-1} p^2\right).
\ee
The function $Q(p,x)$ will still be $x$-dependent in general, as it may contain instances of the fluctuation field $\vp(x)$ and its derivatives. In the next section we will perform a derivative expansion of $Q(p,x)$ which can then be used in the last expression to evaluate both the flow equation and the msWI.

\subsection{Derivative expansion of $Q(p,x)$}
When the ansatz \eqref{equ:ansatzGamma} for the effective action is substituted into the left hand side of the flow equation \eqref{equ:flowGamma-detail} or the msWI \eqref{equ:sWI-detail}, the resulting expression already has the form of a derivative expansion. On their right hand sides however, we encounter the term
\be
\label{equ:Q}
Q(p,x) = e^{-ip\cdot x}\Delta_x e^{i p\cdot x}=e^{-ip\cdot x}\left[\hess+R\right]^{-1}e^{i p\cdot x}
\ee
as it appears in \eqref{equ:traces}. Since it will be more convenient in the following, we have expressed \eqref{equ:Delta} in the more compact and equivalent differential operator notation, temporarily omitting $k$-dependence to avoid clutter. With an inverse of an operator containing the fluctuation field $\vp(x)$ and its derivatives, this term in its current form is not presented in a derivative expansion. In order to be able to compare the left hand sides with the corresponding right hands sides order by order in the derivative expansion we proceed as follows.

First, the definition \eqref{equ:Q} can be rewritten as
\be
\label{equ:temp2}
R\left\{\eipx Q\right\} = \eipx\left(1-\emipx \hess\left\{\eipx Q\right\}\right).
\ee
For the moment assuming that $R$ is an analytic function of its first argument, a derivative expansion of the left hand side can be achieved by using
\begin{multline}
\left(-\partial^2\right)^n\left\{\eipx Q\right\} = \eipx\left[p^{2n}Q -2in p^{2n-2} p^\mu\partial_\mu Q 
	  -n p^{2n-2}\partial^2 Q \right. \\ \left. -2n(n-1)p^{2n-4}p^\mu p^\nu \partial_\mu \partial_\nu Q + \mathcal{O}\big(\partial^3\big)\right]
\end{multline}
in the Taylor expansion
\be
R(-\partial^2,\chi)\left\{\eipx Q\right\} = \sum_{n=0}^\infty \frac{R^{(n)}(0,\chi)}{n!}\left(-\partial^2\right)^n\left\{\eipx Q\right\}.
\ee
Summing this series leaves us with an expansion that makes sense for any $R(-\partial^2,\chi)$:
\begin{multline}
R(-\partial^2,\chi)\left\{\eipx Q\right\} = \eipx \Big[ R(p^2,\chi)Q -2i\partial_{p^2} R(p^2,\chi)p^\mu\partial_\mu Q
		-\partial_{p^2}R(p^2,\chi) \partial^2 Q 
		\\ 
		\left. -2\partial_{p^2}^2 R(p^2,\chi) p^\mu p^\nu \partial_\mu
		\partial_\nu Q + \mathcal{O}\big(\partial^3\big)\right].
\end{multline}
By substituting this result on the left hand side of \eqref{equ:temp2}, we arrive at the following implicit equation for $Q$,
\begin{multline}
\label{equ:Qimp}
Q = R\left(p^2,\chi\right)^{-1}\left[1-\emipx \hess\left\{\eipx Q\right\} +2i\partial_{p^2} R(p^2,\chi)p^\mu\partial_\mu Q \right. \\
    +\left. \partial_{p^2}R(p^2,\chi) \partial^2 Q + 2\partial_{p^2}^2 R(p^2,\chi) p^\mu p^\nu \partial_\mu
		\partial_\nu Q + \mathcal{O}\big(\partial^3\big)\right].
\end{multline}
The Hessian operator associated with the ansatz \eqref{equ:ansatzGamma} can be separated into contributions according to the number of derivatives of $\vp$ they contain,
\be
\label{equ:expHess}
\hess = \hess_0 + \hess_1 + \hess_2
\ee
with the individual terms given by
\be
\hess_0 = V'' + \frac{K}{f}\partial^2, \qquad \hess_1 = \frac{K'}{f}\partial_\mu\vp \partial^\mu, \qquad
	  \hess_2 = \frac{1}{2}\frac{K''}{f}\left(\partial_\mu \vp\right)^2 +  \frac{K'}{f}\partial^2\vp.
\ee
In the same vein, we also expand
\be
\label{equ:Qsum}
Q = Q_0 + Q_1 + Q_2 + \dots,
\ee
collecting terms with $n$ derivatives of $\vp$ into $Q_n$. Derivatives $\partial_\mu Q_n$ therefore are allocated to $Q_{n+1}$ since we infer from the definition \eqref{equ:Q} that $Q$ depends on $x$ only through the fluctuation field $\vp(x)$ and its derivatives as they appear in the Hessian $\hess$ as well as the background field $\chi(x)$ whose derivatives we neglect, however. Using this expansion and \eqref{equ:expHess} in \eqref{equ:Qimp} leads to implicit equations for the first three terms $Q_n, \, n=0,1,2$ that are easily solved to give
\begin{align}
\label{equ:Q0}
Q_0 =& \left[V'' - \frac{K}{f} p^2 + R \right]^{-1}  \\
Q_1 =& -i p^\mu\left[\frac{\partial_\vp K}{f}Q_0 \partial_\mu \vp + 2 \left(\frac{K}{f}+ \partial_{p^2}R\right)\partial_\mu Q_0\right]Q_0 \notag \\
\label{equ:Q2}
Q_2 =& -\left[i p^\mu\left\{\frac{\partial_\vp K}{f}Q_1 \partial_\mu \vp + 2 \left(\frac{K}{f}+ \partial_{p^2}R\right)\partial_\mu Q_1\right\} + \left(\frac{K}{f} + \partial_{p^2}R \right)\partial^2 Q_0
\right. \\ \notag
& \left. + \frac{\partial_\vp K}{f}\left(\partial_\mu \vp \partial^\mu Q_0 + Q_0 \partial^2 \vp\right)
+\frac{1}{2}\frac{\partial_\vp^2 K}{f}Q_0 \left(\partial_\mu \vp\right)^2
+ 2\partial_{p^2}^2R\, p^\mu p^\nu\partial_\mu\partial_\nu Q_0\right]Q_0.
\end{align}
In these expressions $R=R(p^2,\chi)$, and they can now be used directly in \eqref{equ:traces} to obtain both the flow equation \eqref{equ:flowGamma-detail} and the msWI \eqref{equ:sWI-detail}. Out of these $Q_1$ is an odd function of momentum and therefore integrates to zero in \eqref{equ:traces}. This is expected since there are no terms in the effective action with only one derivative of $\vp$, but we display $Q_1$ since it appears in the $\mathcal{O}\big(\partial^2\big)$-term $Q_2$.

\subsection{Evaluation of the flow equation}
We are now in a position to evaluate the flow equation \eqref{equ:flowGamma-detail} for the truncation ansatz \eqref{equ:ansatzGamma} using the result \eqref{equ:traces} with $\mathcal{K}(p^2/f) = \dot R(p^2/f)$. Here and in the following the dot notation refers to a derivative with respect to RG time $t$, as defined in \eqref{time}. Making use of \eqref{equ:Qsum} and \eqref{equ:Q0} we immediately obtain
\be
\label{equ:flowV}
\partial_t V(\vp,\chi) = \frac{\tilde \Omega_{d-1}}{2}f(\chi)^{-\frac{d}{2}}\int dp \,p^{d-1}\frac{\dot R(p^2/f)}{\partial_\vp^2 V - K p^2/f+R(p^2/f)}.
\ee
It involves the constant $\tilde \Omega_{d-1}= \Omega_{d-1}/(2\pi)^d$, where $\Omega_{d-1}$ is the surface area of the $(d-1)$-dimensional sphere, and is expressed as an integral over the modulus $p=|p^\mu|$. The overall background field factor in front of the integral has its origin in the left hand side of the flow equation and comes from the measure factor contained in the ansatz \eqref{equ:ansatzGamma}.

A similar computation at $\mathcal{O}\big(\partial^2\big)$ of the derivative expansion involving \eqref{equ:Q2} leads to
\be
\label{equ:flowK}
f(\chi)^{-1}\partial_t K(\vp,\chi) = \tilde \Omega_{d-1}\, f(\chi)^{-\frac{d}{2}}\int dp\, p^{d-1}\, P\left(p^2,\vp,\chi\right) \dot R\left(p^2/f\right),
\ee
where
\begin{align}
\label{equ:P}
 P =& -\frac{1}{2}\frac{\partial_\vp^2 K}{f}Q_0^2 + \frac{\partial_\vp K}{f}\left(2\partial_\vp^3 V-\frac{2d+1}{d}\frac{\partial_\vp K}{f}p^2\right) Q_0^3 \\ \notag
			  &  -\left[\left\{\frac{4+d}{d}\frac{\partial_\vp K}{f}p^2-\partial_\vp^3V\right\}\left(\partial_{p^2}R-\frac{K}{f}\right) + \frac{2}{d}p^2\partial_{p^2}^2 R\left(\frac{\partial_\vp K}{f}p^2-\partial_\vp^3 V\right)\right]\left(\partial_\vp^3 V -\frac{\partial_\vp K}{f}p^2\right) Q_0^4 \\ \notag
 &-\frac{4}{d}p^2\left(\partial_{p^2} R-\frac{K}{f}\right)^2\left(\partial_\vp^3V-\frac{\partial_\vp K}{f}p^2\right)^2 Q_0^5
\end{align}
is obtained from \eqref{equ:Q2} by integration by parts under the integral in \eqref{equ:flowGamma-detail}. We have chosen to arrange this expression by orders of powers of the $\mathcal{O}\big(\partial^0\big)$-term $Q_0$, \eqref{equ:Q0}.

The flow equations \eqref{equ:flowV} and \eqref{equ:flowK} are still rather general as we have not specified the form of the cutoff operator $R(z)$. We will come back to this in sec.  \ref{sec:independence}.

\subsection{Evaluation of the modified split Ward identity}
\label{sec:evalsWI}
In principle the msWI \eqref{equ:sWI-detail} can be obtained along the same lines as the flow equation in the previous section. Due to the type of truncation we made for the effective action, \eqref{equ:ansatzGamma}, there is however an additional small but important subtlety that needs to be taken care of. As emphasised just before sec. \ref{sec:strategy-traces} we consider slowly varying background fields $\chi(x)$ so that we can take functional derivatives as in the left hand side of the msWI \eqref{equ:sWI-detail} but neglect all derivatives $\partial_\mu \chi$ compared to derivatives of the fluctuation field $\partial_\mu \vp$, in accordance with the truncation ansatz \eqref{equ:ansatzGamma}. While the derivation of the msWI in sec. \ref{sec:sWI-derivation} is completely general, we have to neglect derivatives $\partial_\mu \eps(x)$ of the shifts in \eqref{equ:split-symmetry} for our truncation of the effective action in order to remain in the regime of slowly varying background fields $\chi(x)$. To do this it is convenient to start from the version \eqref{equ:sWI-detail-eps} of the msWI as it explicitly contains the shifts $\eps$. With the ansatz \eqref{equ:ansatzGamma} for the effective action, the integrand on its left hand side becomes
\be
\label{equ:lhs-sWI}
f^{-\frac{d}{2}}\left(\frac{\delta \Gamma_k}{\delta \chi}-\frac{\delta \Gamma_k}{\delta \vp}\right)
=-\frac{1}{2 f}\left\{\partial_\chi K - \partial_\vp K +\frac{d-2}{2}\dclnf K\right\}
\left(\partial_\mu \vp\right)^2+\frac{d}{2}\dclnf V+\partial_\chi V- \partial_\vp V + I,
\ee
where, writing $K(x) = K(\vp(x),\chi(x))$, the last term is
\be
\label{equ:I}
I(w) = f(w)^{-\frac{d}{2}}\int d^dx f(x)^\frac{d}{2}\frac{K(x)}{f(x)}\partial_\mu \delta(x-w)\partial^\mu \vp(x).
\ee
Hence, for slowly varying $\eps$ we have
\be
\int_w I(w)\,\eps(w) = 0,
\ee
causing this term to drop out on the left hand side of the msWI \eqref{equ:sWI-detail-eps}. We note here that we would arrive at the same result for the left hand side of \eqref{equ:sWI-detail} if we consider the delta function in \eqref{equ:I} to be slowly varying itself, $\partial_\mu \delta(x-w) = 0$.

The right hand side of the msWI \eqref{equ:sWI-detail-eps} can be evaluated using the techniques of sec. \ref{sec:strategy-traces}. Instead of \eqref{equ:KsWI} we use 
\be
\label{equ:op-sWI}
\mathcal{K}_x = \eps(x)\left\{\frac{d}{2}\,\partial_\chi\! \ln\! f(x) \, R_{k,x} + \partial_\chi R_{k,x}\right\}
\ee
resulting in the appropriately modified version of \eqref{equ:traces},
\be
\label{equ:rhs-sWI}
T_\mathcal{K} = \int d^dx \frac{d^dp}{(2\pi)^d}\,Q(p,x)\left[\frac{d}{2}\dclnf R_k(p^2/f) + \partial_\chi R_k(p^2/f)\right]\eps(x).
\ee
At order $\mathcal{O}\big(\partial^0\big)$ of the derivative expansion \eqref{equ:Qsum} with \eqref{equ:Q0}, this last expression can be used directly with the corresponding part of \eqref{equ:lhs-sWI} in \eqref{equ:sWI-detail-eps} to get the msWI for the potential,
\be 
\label{equ:sWI-V}
\partial_\chi V- \partial_\vp V+\frac{d}{2}\dclnf V
= \frac{\tilde \Omega_{d-1}}{2}f(\chi)^{-\frac{d}{2}}\int dp\, p^{d-1} \,\frac{\partial_\chi R(p^2/f)+\frac{d}{2}\dclnf R(p^2/f)}{\partial^2_\vp V -Kp^2/f +R(p^2/f)}.
\ee
The last term on the left hand side stems from the measure factor in the ansatz \eqref{equ:ansatzGamma} for the effective action. In the same way,  the second term in the numerator on the right hand side has its origin in the measure factor of the cutoff action \eqref{equ:cutoff-action}, whereas the first term is a result of the fact that the cutoff is imposed with respect to the eigenmodes of the background field Laplacian, $-\bar \nabla^2$.

In order to evaluate \eqref{equ:rhs-sWI} at second order of the derivative expansion using \eqref{equ:Q2}, we integrate by parts under the spatial integral in \eqref{equ:rhs-sWI}, just as for the flow equation. The advantage of the representation \eqref{equ:rhs-sWI} is that it makes it apparent that this integration by parts does not pick up additional terms from the operator \eqref{equ:op-sWI} as long as we neglect all derivatives of the background field and the shift function $\eps(x)$. Thus, collecting the $\left(\partial_\mu \vp\right)^2$-terms from \eqref{equ:lhs-sWI} for the left hand side of the msWI \eqref{equ:sWI-detail-eps}, the msWI for the kinetic term assumes the form
\be 
\label{equ:sWI-K}
\frac{1}{f}\left\{\partial_\chi K - \partial_\vp K +\frac{d-2}{2}\dclnf K\right\}
= \tilde \Omega_{d-1}f^{-\frac{d}{2}}\int dp\, p^{d-1} \,P(p^2,\vp,\chi)\left[\partial_\chi R+\frac{d}{2}\dclnf R\right].
\ee
Here, $P$ is the same as for the flow equation, \eqref{equ:P}, and from now on the argument of the cutoff is understood to be $R=R(p^2/f)$. On the right hand side we find the two contributions analogous to the right hand side of the msWI for the potential, \eqref{equ:sWI-V}, and on the left hand side the factor $(d-2)/2 = d/2 -1$ comprises the contribution coming from the measure factor and from the background covariant derivative of the kinetic term in the ansatz \eqref{equ:ansatzGamma}.

We also note that for the right hand side of \eqref{equ:sWI-K} too, we could have instead started from \eqref{equ:sWI-detail} with \eqref{equ:KsWI} as long as we regard the delta function $\delta(x-w)$ to be slowly varying, allowing us to neglect its derivatives. However, the justification for this can only be understood when using the form \eqref{equ:sWI-detail-eps} of the msWI together with a correct treatment of the shift function $\eps(x)$ as above.

\section{Combining the msWI and flow equations}
\label{sec:combine}

In this section we combine the flow equation and msWI, resulting in flow equations with respect to a new scale $\hat k$ that are explicitly independent of the background field $\chi$ and moreover explicitly independent of $f$. Then we compare fixed points under these flows to fixed points under the original flow equations.

\subsection{Background dependence}
\label{sec:dependence}

Our starting point is the collection of flow and msWI equations for $V$ and $K$, as given by \eqref{equ:flowV}, \eqref{equ:flowK}, \eqref{equ:sWI-V} and \eqref{equ:sWI-K}, where $P$ defined  through \eqref{equ:P} using  \eqref{equ:Q0}. The reader can easily verify that the remnant background diffeomorphism invariance \eqref{remnant} is respected by this system, since through the Fourier transform relations, \eqref{remnant} implies also
\be
\label{p-diffeo}
p\mapsto \lambda\, p\,.
\ee

Considering the flow equations for the potential \eqref{equ:flowV} and for the kinetic term \eqref{equ:flowK}, it would seem that the background field enters through the parametrisation function $f$ of \eqref{equ:confpar}. As discussed previously this is certainly not surprising in the present context of the conformal factor of quantum gravity. Note however that there appear no derivatives with respect to the background field in either of the flow equations, a direct consequence of the general structure of the flow equation of the effective action \eqref{equ:flowGamma}. This means that actually the flow equations are then independent of $f$, as can be seen by choosing $\lambda=\sqrt{f(\chi)}$ and making the change of variables \eqref{p-diffeo} in the integrals on their right hand sides. The flow equations themselves therefore also have no explicit dependence on $\chi$ or its derivatives.

Na\"\i vely we can therefore arrive at background independent solutions 
\be
\nonumber
{\cal S}_0 = \{V(\vp,\chi) \equiv V_t(\vp),\, K(\vp,\chi) \equiv K_t(\vp)\}
\ee 
by simply declaring  that $K$ and $V$ are independent of $\chi$, as shown. But such solutions clearly destroy the idea that the fundamental underlying theory should depend on the total field $\phi=\vp+\chi$. 
Alternatively for the flow equations we can just regard $\chi$ as an auxiliary parameter and use it to label a range of different solutions in terms of $\vp$ and $t$:
\be
\nonumber
{\cal S}_\chi = \left\{V_t(\vp,\chi), K_t(\vp,\chi)\right\}\,.
\ee 
Adapting the simple example discussed in the Introduction, we note that if $V_t(\vp,\chi)$ is a solution then $V_t(\vp,\chi)+F(\chi)$ is also a solution of \eqref{equ:flowV} and \eqref{equ:flowK}, where $F$ is an arbitrary $t$-independent function of the background field. We see now that the freedom is far larger. Apart from a tacit assumption of continuity in $\chi$, there is complete freedom in how we choose to scan through different solutions, \ie implement dependence on $\chi$. 

All this illustrates that from an investigation of the flow equation(s) alone we cannot control the background field dependence of the renormalization group solutions sufficiently. Indeed it is then by no means possible to guess what background independence is meant to mean, let alone ensure that it is established. This makes concrete the discussion we set out in the Introduction. The fact that it is possible to successfully quantise for all background configurations simultaneously does not guarantee background independence. Instead the result is simply a separate effective action for each background field configuration.

Therefore, what seems to be called for is an additional identity that complements the flow equation in a way that involves the background field as an independent variable in a differential equation, rather than just as a parameter. This is the r\^ole played by the broken split Ward identity given by \eqref{equ:sWI-V} for the potential and by \eqref{equ:sWI-K} for the kinetic term. The expectation will be that the msWI, when used in conjunction with the flow equation, will eliminate the additional freedom represented by the background field as described here. One aim of this section is to show that this can be achieved by appropriately combining the msWI with the flow equation.

Note that the msWI equations  \eqref{equ:sWI-V} and \eqref{equ:sWI-K} do depend on $f(\chi)$ in a non-trivial way. In particular making the change of variables \eqref{p-diffeo} with $\lambda=\sqrt{f(\chi)}$, does not eliminate $f(\chi)$ from these equations. Nevertheless when appropriately combined with the flow equations, we will again find that the parametrisation function $f$ drops out.

\subsection{Background independence}
\label{sec:independence}

We begin by redefining
\begin{equation}
 V \mapsto \frac{\tilde \Omega_{d-1}}{2}V, \qquad K \mapsto \frac{\tilde \Omega_{d-1}}{2}K, \qquad R\mapsto \frac{\tilde \Omega_{d-1}}{2}R
\end{equation}
to convert the flow equation \eqref{equ:flowV} and the msWI \eqref{equ:sWI-V} for the potential into
\begin{subequations}
\label{equ:sysV-resc}
\begin{align}
 \partial_t V(\vp,\chi) &= f(\chi)^{-\frac{d}{2}}\int dp \,p^{d-1}\,Q_0\,\dot R, \label{equ:flowV-resc}\\
 \partial_\chi V- \partial_\vp V+\frac{d}{2}\dclnf V
&= f(\chi)^{-\frac{d}{2}}\int dp\, p^{d-1} \,Q_0\left[\partial_\chi R+\frac{d}{2}\dclnf R\right], \label{equ:sWIV-resc}
\end{align}
\end{subequations}
where we have made use of the zeroth order of the derivative expansion $Q_0$ given in \eqref{equ:Q0}. Rescaling the cutoff operator by a positive constant as we have done here is allowed since it does not affect the suppression of modes it is responsible for. In the same way, the flow equation \eqref{equ:flowK} and the msWI \eqref{equ:sWI-K} for the kinetic term are transformed into
\begin{subequations}
\label{equ:sysK-resc}
\begin{align}
 f^{-1}\partial_t K(\vp,\chi) &= 2 f^{-\frac{d}{2}}\int dp\, p^{d-1}\, P\left(p^2,\vp,\chi\right) \dot R, \label{equ:flowK-resc}\\
 f^{-1}\left\{\partial_\chi K - \partial_\vp K +\frac{d-2}{2}\dclnf K\right\}
&= 2f^{-\frac{d}{2}}\int dp\, p^{d-1} \,P(p^2,\vp,\chi)\left[\partial_\chi R+\frac{d}{2}\dclnf R\right]\,, \label{equ:sWIK-resc}
\end{align}
\end{subequations}
where $P$ is still given by \eqref{equ:P}.

So far we have left the functional form of $R_k(p^2/f)$ completely general. In fact it is constrained, in particular by the need to respect scaling dimensions. As intimated in the Introduction, there are a number of subtleties in implementing such dimensions in the present context. We will carefully reason through these subtleties in sec. \ref{sec:dimensions}. For now we just make the minimum justifications and accept the assignments as intuitively reasonable, since we want to demonstrate that the mathematical structure alone is enough to be guided towards a deeper description in terms of a new background independent scale $\hat k$ and background invariant flow equations. 

We know that we have to allow generically for an anomalous dimension of the conformal fluctuation field $[\vp]=\eta/2$. Then in order for the left hand sides of the broken split Ward identities \eqref{equ:sWIV-resc} and \eqref{equ:sWIK-resc} to be consistent in terms of dimensions, the background field also has to have $[\chi]=\eta/2$. We can allow for a general dimension for $f$ also, setting $[f]=d_f$. In particular, as discussed in the Introduction, it will be convenient to entertain the possibility that $d_f\ne0$. 
Due to its place in the cutoff action \eqref{equ:cutoff-action} we thus write \cite{Morris:1994ie}:
\begin{equation}
\label{equ:cutoff}
 R\left(p^2/f\right) = - k^{d-\eta-\frac{d}{2}d_f}\,r\!\left(\frac{p^2}{k^{2-d_f}f}\right)\,,
\end{equation}
where the profile function $r(z)$ is a dimensionless function of a dimensionless argument. The crucial ingredient here is the minus sign on the right hand side, reflecting the minus sign of the conformal factor in quantum gravity as mentioned in sec. \ref{sec:eff-ac}. At a practical level, this sign ensures that singularities are avoided in the modified propagator $\left[\Gamma_k^{(2)}+R_k\right]^{-1}$ in \eqref{equ:flowGamma}, \cf \cite{Reuter:1996}.
The cutoff operator performs the suppression of modes correctly if the profile function $r(z)$ satisfies the requirements $\lim_{z\to0} r(z) >0$ and $\lim_{z\to\infty} r(z) = 0$, see \eg \cite{Morris:1994ie,Litim:2001}.

Now we come to the crux of the matter. How can we combine the systems of flow and msWI equations \eqref{equ:sysV-resc}, \eqref{equ:sysK-resc}, to reach a deeper background independent description? We are inspired to ask this question by the dramatic demonstration that background independence can be restored by a certain change of variables \cite{Bridle:2013sra}. However in ref. \cite{Bridle:2013sra}, only LPA for scalar field theory was treated, and this change of variables was guessed at, using physical intuition. In the current case it is not so easy to guess the result. Indeed {\it a priori} it is not even any longer clear  that a background independent RG description should exist.

In appendix \ref{app:scalar} we show that we could have deduced the change in variables in ref. \cite{Bridle:2013sra} directly from the structure of the flow and msWI equations by recognising that they can be combined by eliminating their right hand sides, leaving a linear partial differential equation, which can then be solved by the method of characteristics.  This provides the change of variables which in turn collapses the equations to manifestly background independent form.

We can generalise this technique to the current situation, at the price of further restrictions on  $R$ when $\eta\ne0$. Concentrating on the right hand sides of the $V$ equations \eqref{equ:sysV-resc}, we see that they differ only in the final term, namely $\dot R$ for the flow equation, and $\left[\partial_\chi R+\frac{d}{2}\dclnf R\right]$ for the msWI. Concentrating on the right hand sides of the $K$ equations \eqref{equ:sysK-resc}, we see that they also only differ in the final term and furthermore in the same way. Therefore if we can arrange 
for these final terms to have the same $p$ dependence, the right hand sides of the $V$ equations become essentially the same, and likewise the $K$ equations. This allows the flow and msWI equations to be combined into linear partial differential equations, which again we solve to deduce the required changes of variable. 

From \eqref{equ:cutoff}, we see that
\be
\dclnf {\dot R} = (2-d_f) \left[\partial_\chi R+\frac{d}{2}\dclnf R\right] -\eta \dclnf R\,.
\ee
Since $\dclnf$ does not depend on $p$, we see we automatically achieve our aim if $\eta=0$, without further restriction on $R$. However if $\eta\ne0$, then to achieve our aim we need also
\be 
\eta \dclnf R = A  \left[\partial_\chi R+\frac{d}{2}\dclnf R\right]\,,
\ee
for some constant $A$. Rearranging, we see that this implies
\be 
f {\partial R\over\partial f} = \left( {\eta\over A} - \frac{d}{2}\right) R\,,
\ee
\ie, recalling \eqref{equ:cutoff}, that $R$ is a power of its argument. Even though we do not need the restriction in the $\eta=0$ case, we can use such a power-law cutoff for this case too. These types of cutoff have the added benefit that they ensure that the derivative expansion approximation preserve the quantisation of the anomalous dimension in non-gravitational systems, \eg scalar field theory \cite{Morris:1994ie,Morris:1994jc,Morris:1998}.
Hence, we set
\begin{equation}
 \label{equ:cutoff-pwrlw}
 r(z) = \frac{1}{z^n}, \qquad \text{implying} \qquad zr'(z) = -nr(z).
\end{equation}
To ensure finiteness of the integrals on the right hand sides of \eqref{equ:sysV-resc} and \eqref{equ:sysK-resc}, the exponent $n$ has to be chosen such that $n>d/2-1$, \cf \cite{Morris:1994ie}. From \eqref{equ:cutoff} we also need to ensure that
\be
\label{bad-n}
n\ne\frac{\eta}{2-d_f}-\frac{d}{2}\,,
\ee
otherwise $R$ becomes independent of $k$.
In terms of such an $r$ we then have the replacements
\begin{subequations}
\begin{align}
 \partial_t R &= - k^{d-\eta-\frac{d}{2}d_f}\left(d-\eta+2n-\frac{d+2n}{2}d_f\right)r\!\left(\frac{p^2}{k^{2-d_f}f}\right),  \\
 \partial_\chi R+\frac{d}{2}\dclnf R &= - k^{d-\eta-\frac{d}{2}d_f}\frac{d+2n}{2}\dclnf r\!\left(\frac{p^2}{k^{2-d_f}f}\right).
\end{align}
\end{subequations}
Exploiting this allows us to eliminate the right hand sides, combining the flow equation \eqref{equ:flowV-resc} with the msWI \eqref{equ:sWIV-resc} into a single equation,
\begin{equation}
\label{equ:combV}
 \dclnf \partial_t V = \alpha \left(\frac{d}{2}\dclnf V + \partial_\chi V-\partial_\vp V\right),
\end{equation}
where the constant $\alpha$ is given by
\begin{equation}
\label{eq:alpha}
 \alpha = 
 2\left(1-\frac{\eta}{d+2n}\right)-d_f\,,\qquad\alpha\ne0\,,
\end{equation}
and where the inequality follows from \eqref{bad-n}.
The flow equation for the kinetic term \eqref{equ:flowK-resc} and its msWI \eqref{equ:sWIK-resc} can also be combined to give the similar equation
\begin{equation}
\label{equ:combK}
 \dclnf \partial_t K = \alpha\left(\partial_\chi K - \partial_\vp K +\frac{d-2}{2}\dclnf K\right).
\end{equation}
The two combined equations \eqref{equ:combV} and \eqref{equ:combK} are solved by the method of characteristics to give the change of variables
\begin{equation}
 \label{equ:chvars}
 V(k,\vp,\chi) = f(\chi)^{-\frac{d}{2}}\tilde V(\tilde k,\phi), \qquad K(k,\vp,\chi) = f(\chi)^{-\frac{d}{2}+1}\tilde K(\tilde k,\phi), \qquad \tilde k = k f(\chi)^\frac{1}{\alpha},
\end{equation}
where we have highlighted the dependence on the original and the transformed RG scale and used the total field $\phi = \chi + \vp$. Under this change of variables the systems \eqref{equ:sysV-resc} for the potential and \eqref{equ:sysK-resc} for the kinetic term are respectively equivalent to the first and second equation in the following system,
\begin{subequations}
 \label{equ:sys-red}
 \begin{align}
  \partial_{\tilde t}\tilde V &= -\frac{1}{2}(2n+d)\alpha\,\tilde k^{\alpha(n+d/2)}\int dp\, p^{d-1}\, \tilde Q_0\,r\!\left(p^2\right),   \\
  \partial_{\tilde t}\tilde K &= -(2n+d)\alpha\,\tilde k^{\alpha(n+d/2)}\int dp\, p^{d-1}\, \tilde P\,r\!\left(p^2\right).
 \end{align}
\end{subequations}
Here $\partial_{\tilde t} \equiv \tilde k \partial_{\tilde k}$, and $\tilde Q_0$ and $\tilde P$ are given by applying the change of variables \eqref{equ:chvars} to \eqref{equ:Q0} and \eqref{equ:P} and pulling out a factor of $f^{d/2}$ (temporarily suppressing the $\chi$ dependence). Equivalently in  \eqref{equ:Q0} and \eqref{equ:P}, replace $V$ by $\tilde V$, $K/f$ by $\tilde K$, and $R$ by $f^{d/2}R$. Since it turns out from \eqref{equ:cutoff}, \eqref{equ:cutoff-pwrlw} and \eqref{equ:chvars}, that 
\be
\label{R-ktilde}
f^\frac{d}{2}R(p^2/f) = -\,\tilde{k}^{\frac{1}{2}(d+2n)\alpha} / p^{2n}\,,
\ee
we see that the change of variables indeed has removed all dependence on both $\chi$ and $f$ from the new system of equations \eqref{equ:sys-red}.

The next step consists in adopting dimensionless variables.  We remark that the mass dimension of the transformed RG scale $\tilde k$ may deviate from one, $\big[\tilde k    \big]=1+d_f/\alpha$. It is therefore helpful to define
a new RG scale given by
\begin{equation}
\label{equ:kbar}
 \kh = \tilde k^\frac{\alpha}{\alpha+d_f}\,,
\end{equation}
in terms of which the dimensionless versions of the flow equations take their simplest and most natural form.
The mass dimension of this is $\big[\kh\big]=1$ and the dimensionless quantities are then given by the following redefinitions,
\begin{equation}
\label{equ:dimless}
 \tilde V = \kh^{d}\,\Vh, \qquad \tilde K = \kh^{d-2-\eta}\Kh, \qquad \phi = \kh^{\frac{\eta}{2}}\ph, \qquad p = \kh\,\hat p\,.
\end{equation}
Note that if $\big[\tilde k    \big]=0$ we cannot make variables dimensionless by using $\tilde{k}$, and the definition of $\hat k$ then makes no sense. Since making the variables dimensionless is equivalent to the blocking step in the Wilsonian RG framework \cite{Morris:1994ie,Morris:1998}, in this case the Wilsonian RG framework breaks down. We therefore need to utilise the freedom to choose the cutoff exponent $n$ so that
\be
\label{bad-n2}
\alpha\ne-d_f\qquad\hbox{or equivalently}\qquad\eta \ne d+2n\,,
\ee
using \eqref{eq:alpha}.
Combining \eqref{equ:kbar} and the last equation in \eqref{equ:chvars} we obtain \eqref{khat}, the already advertised expression for $\hat k$.

Given that the momentum dependence of the cutoff $R$ in \eqref{equ:cutoff}
is held in the cutoff profile $-r(z)$, and 
that $f(\chi)$ has disappeared from the equations, since $[\hat k]=[p]=1$, $[x]=-1$ and $[\vp]=\eta/2$, by dimensions  it has to be the case that 
\begin{equation}
\label{cutoff-kbar}
f^{\frac{d}{2}}\, R\big(p^2/f\big) = - \, \kh^{d-\eta}\, r\big(p^2/\kh^2\big)\,,
\end{equation}
as can also be readily verified from \eqref{R-ktilde} and \eqref{equ:kbar}. The transformation thus
provides a RG scale that converts the cutoff function into the form expected in scalar field theories, \eg \cite{Morris:1994ie,Litim:2002cf}, except for the sign reflecting the conformal factor problem. We will pursue this connection further in sec. \ref{sec:comp-sft}.

Expressing the system \eqref{equ:sys-red} in terms of the dimensionless variables then finally leads to
\begin{subequations}
\label{equ:sys-red-final}
\begin{align}
 \partial_{\hat t} \Vh +d\Vh-\frac{\eta}{2}\ph \Vh' &= -\left(d-\eta+2n\right)\int d\hat p\, \hat p^{d-1} \,\hat Q_0 \,r\big(\hat p^2\big), \label{equ:flowV-final} \\
 \partial_{\hat t} \Kh +(d-2-\eta)\Kh-\frac{\eta}{2}\ph \Kh' &= -2(d-\eta+2n)\int d\hat p\, \hat p^{d-1} \,\hat P\big(\hat p^2,\ph\big) \,r\big(\hat p^2\big). \label{equ:flowK-final}
\end{align}
\end{subequations}
where $\hat t = \ln (\hat k/\mu)$ and where now
\begin{align}
 \hat Q_0 = &\left[\Vh''-\Kh \hat p^2-r\big(\hat p^2\big)\right]^{-1} \label{Q0final} \\ 
 \hat P = &-\frac{1}{2}\Kh'' \hat Q_0^2 + \Kh'\left(2\Vh'''-\frac{2d+1}{d}\Kh' 
 			\hat p^2\right)\hat Q_0^3    \label{Pfinal} \\
    & +\left[\left\{\frac{4+d}{d}\Kh' \hat p^2 - \Vh'''\right\}\left(r'(\hat p^2)+\Kh\right)+\frac{2}{d}\hat p^2 r''\big(\hat p^2\big)\left(\Kh' \hat p^2-\Vh'''\right)\right]\!\left(\Vh'''-\Kh' \hat p^2\right)\!\hat Q_0^4
     \notag \\
    & -\frac{4}{d}\hat p^2 \left(r'\big(\hat p^2\big)+\Kh\right)^2 \left(\Vh'''-\Kh' \hat p^2\right)^2 \hat Q_0^5 \, , \notag 
\end{align}
and primes denote derivatives with respect to $\ph$. We will refer to the two equations \eqref{equ:sys-red-final} as the \emph{reduced system}. It consists of two partial differential equations for $V_{\kh}(\ph)$ and $\Kh_{\kh}(\ph)$ with respect to the RG scale $\kh$ and the total conformal factor field $\ph$.

The solutions of the original systems \eqref{equ:sysV-resc} and \eqref{equ:sysK-resc} comprising the flow equation and msWI for the potential and the kinetic term are mapped onto the solutions of the reduced system \eqref{equ:sys-red-final} in a one-to-one fashion by the changes of variables \eqref{equ:chvars}, \eqref{equ:kbar} and \eqref{equ:dimless}. In this way, the RG flow of the ansatz \eqref{equ:ansatzGamma} for the effective action is fully captured by a set of differential equations that do no longer depend on the background field $\chi$ or the parametrisation function $f$ in \eqref{equ:confpar}.   Therefore, the existence of fixed points, the structure of their eigenspectra, as well as the presence of possible global RG trajectories are not affected by the way the conformal factor is parametrised in \eqref{equ:confpar}.
 {The msWI \eqref{equ:sWiGamma} has made it possible to gain complete control over the dependence of the potential and the kinetic term on the background field and has led to the significant simplifcation as embodied by the reduced system \eqref{equ:sys-red-final} compared to the original systems \eqref{equ:sysV-resc}, \eqref{equ:sysK-resc}.}

The change of variables \eqref{equ:chvars} together with \eqref{equ:kbar} turns the effective action \eqref{equ:ansatzGamma} into
\begin{equation}
\label{equ:Gamma-hat}
 \Gamma_k[\vp,\chi] = \int d^dx \left(-\frac{1}{2} \tilde K_{\kh}(\phi) \left(\partial_\mu \phi\right)^2+ \tilde V_{\kh}(\phi)\right),
\end{equation}
where the two RG scales are related by \eqref{khat}, and we have made use of the approximation of slow background fields $\partial_\mu \chi =0$. Hence the solutions are such that the measure factor is absent in \eqref{equ:Gamma-hat} but the right hand side still depends on the background field through $\kh$. This is expected since the right hand side of the msWI \eqref{equ:sWiGamma} contains the background field dependent cutoff operator which is an intrinsic feature of the RG flow. As mentioned in the Introduction, the cutoff operator drops out in the limit $k\to0$ and the effective action should become a function of the total field $\phi$ only. As we can see from \eqref{khat}, this is indeed the case. If the exponent of $k$ is positive the limit $k\to 0$ implies $\kh \to 0$ and the right hand side of \eqref{equ:Gamma-hat} becomes a function of $\phi$ only. This remains true if the exponent of $k$ in \eqref{khat} is negative, where now $k\to 0$ corresponds to $\kh \to \infty$ and the right hand side of \eqref{equ:Gamma-hat} would represent an ultraviolet fixed point of \eqref{equ:sys-red-final}. In either case this leads to a background independent effective action in the limit $k\to 0$ as a result of combining the msWI with the flow equation.

\subsection{Equivalence of RG flows and (non-)existence of $k$-fixed points}
\label{sec:fpmap}
As emphasised in the previous section, the reduced system \eqref{equ:sys-red-final} is equivalent to the original system of flow equations \eqref{equ:flowV} and \eqref{equ:flowK} when also the equations of the msWI \eqref{equ:sWI-V} and \eqref{equ:sWI-K} are imposed. However, this is an equivalence between differential equations. By contrast, we now ask if the RG flow described by the reduced system \eqref{equ:sys-red-final} expressed in terms of the background independent RG scale $\kh$ is equivalent to the RG flow of the original equations \eqref{equ:flowV} and \eqref{equ:flowK}. { Such an equivalence of RG flows would require in particular that fixed points of one system correspond to fixed points of the other, and we will see now that whether this is true in the present case depends on the parametrisation function $f$.} Furthermore we will discover that in the general case, the inherent dependence on the background field $\chi$, forced by the background dependence of the cutoff (and policed by the msWI), actually forbids the very existence of $k$-fixed points.

Let us first concentrate on fixed point solutions for the potential. Denoting quantities that have been made dimensionless using the original RG scale $k$, with a bar, we write
\begin{equation}\label{Vh}
V(\vp,\chi) = k^{d-\frac{d}{2}d_f} \bar V(\bp, \cb), \qquad \vp = k^{\eta/2}\bp, \qquad \chi = k^{\eta/2}\cb
\end{equation}
for the original potential of \eqref{equ:flowV}, resulting in fixed point solutions satisfying $\partial_t \bar V =0$. On the other hand we have from the change of variables \eqref{equ:chvars} and the corresponding relation in \eqref{equ:dimless} that
\begin{equation}\label{Vt}
V(\vp,\chi) = \kh^d f^{-d/2} \Vh(\ph), \qquad \ph = \left(k/\kh\right)^{\eta/2}(\cb+\bp) \,.
\end{equation}
Using the relation \eqref{khat} between the two RG scales, the last two equations combine to give
\begin{equation}
\bar V(k,\bp,\cb) = \left(k^{-d_f} f\right)^\frac{d\eta}{2(d-\eta+2n)} \Vh(\kh,\ph).
\end{equation}
Recalling \eqref{time} and bearing in mind that the partial derivative in the fixed point condition $\partial_t \bar V=0$ does not act on the dimensionless arguments $\bp$ and $\cb$, the last relation leads to
\begin{align}
\label{fpcondV}
\partial_t \bar V = \left(k^{-d_f} f\right)^\frac{d\eta}{2(d-\eta+2n)} & \left\{ \frac{1}{\al +d_f}
				\left(\al+ \frac{\eta}{2}\chi \dclnf\right) \partial_{\hat t} \Vh \right. \notag \\
				&\left.
				+ \left(\frac{\eta}{2}\chi \dclnf -d_f\right)\left(\frac{d\eta}{2(d-\eta+2n)}\Vh
				-\frac{\eta}{2(\al+d_f)}\ph \Vh'\right)\right\}\,.
\end{align}
The kinetic term $K$ can be dealt with similarly. Changing to the dimensionless version using $k$ gives the analogue of \eqref{Vh},
\begin{equation}
\label{Kh}
K(\vp,\chi) = k^{d-2-\eta-\left(\frac{d}{2}-1\right)d_f} \bar K(\bp,\cb),
\end{equation}
and the analogue of \eqref{Vt} is
\begin{equation}
K(\vp,\chi) = \kh^{d-2-\eta}f^{\frac{d}{2}+1} \Kh(\ph).
\end{equation}
Combining these results in
\begin{equation}
\bar K(\bp,\cb) = \left(k^{-d_f}f\right)^{-\frac{(n+1)\eta}{d-\eta+2n}} \Kh(\ph)
\end{equation}
which we use to calculate
\begin{align}
\label{fpcondK}
\partial_t \bar K = \left(k^{-d_f} f\right)^{-\frac{(n+1)\eta}{d-\eta+2n}} & \left\{ \frac{1}{\al +d_f}
				\left(\al+ \frac{\eta}{2}\chi \dclnf\right) \partial_{\hat t} \Kh \right. \notag \\
				&\left.
				- \left(\frac{\eta}{2}\chi \dclnf -d_f\right)\left(\frac{(n+1)\eta}{d-\eta+2n}\Kh
				+\frac{\eta}{2(\al+d_f)}\ph \Kh'\right)\right\}.
\end{align}
From \eqref{fpcondV} and the last equation we see that in general the $\hat k$-fixed point condition $\partial_{\hat t} \Vh=\partial_{\hat t} \Kh=0$ of the reduced flow \eqref{equ:sys-red-final} does not imply $k$-fixed points $\partial_t \bar V = \partial_t \bar K =0$ of the original flow. 

In fact for flows in $k$, the situation is in general dramatically worse than that. Since solutions $\Vh(\ph)$ and $\Kh(\ph)$ are in one-to-one correspondence with solutions $\bar V(\bp,\cb)$ and $\bar K(\bp,\cb)$ of the original system, but depend on one less variable, equations \eqref{fpcondV} and \eqref{fpcondK} actually imply in general that $k$-fixed points cannot exist. To see this note that holding $\hat t$ fixed implies through \eqref{khat} that $k$ varies as $f(\chi)^{-1/\alpha}$. Through \eqref{Vh}, this then also fixes the dependence of $\cb$ in terms of $\chi$. From \eqref{equ:dimless}, holding $\ph$ fixed then fixes also the dependence of $\bp$ in terms of $\chi$. With these dependences in place we can still freely vary $\chi$ while holding all background independent (\ie hatted) quantities fixed. 
Assuming $\eta\dclnf$ varies with $\chi$, and noting \eqref{bad-n2},    we thus see from \eqref{fpcondV} that the fixed point condition $\partial_t \bar V =0$ implies both $\partial_{\hat t} \Vh =0$ and 
\be
\frac{d\eta}{2(d-\eta+2n)}\Vh-\frac{\eta}{2(\al+d_f)}\ph \Vh' = 0\qquad\implies\qquad \Vh\propto \ph^{2d\over d+2n}\,,
\ee
which by inspection of \eqref{equ:flowV-final}, is impossible. With the same assumption and reasoning, we see from \eqref{fpcondK} and \eqref{equ:flowK-final} that $\partial_{ t} \bar K=0$ is also impossible. Therefore we have shown that if $\eta\dclnf$ varies with $\chi$,  
fixed points with respect to $k$ cannot exist. 

If $\eta\dclnf$ does not depend on $\chi$ then one of three possibilities must be met:
\begin{enumerate}
\item $\eta\ne0$ and $f(\chi)\propto\chi^\gamma$ for some power $\gamma\ne 2d_f/\eta$,
\item $\eta\ne0$ and $f(\chi)\propto\chi^{2d_f/\eta}$,
\item $\eta=0$ ($f$ of any form and $d_f$ any value).
\end{enumerate}
In the first case $\partial_t \bar V =0$ implies 
\be
\label{poss3}
\partial_{\hat t} \Vh + A\left\{\frac{d\eta}{2(d-\eta+2n)}\Vh-\frac{\eta}{2(\al+d_f)}\ph \Vh'\right\} = 0\,,
\ee
for some constant $A\ne0$. Therefore substituting this equation into \eqref{equ:flowV-final} results in a non-linear ordinary differential equation which potentially has solutions $\Vh(\ph)$ but which, by \eqref{poss3}, do not correspond to a $\hat k$-fixed point, $\partial_{\hat t} \Vh=0$. The same arguments establish the corresponding result for the kinetic term $K$. Therefore we have shown that in this case $k$-fixed points can exist but then do not correspond to $\hat k$-fixed points.

In the final two cases the second term in braces vanishes, in both \eqref{fpcondV} and \eqref{fpcondK}, thus implying that in these cases the two fixed point conditions are now equivalent:
\begin{equation}
\label{coincide}
\partial_t \bar V = \partial_t \bar K =0 \qquad \Leftrightarrow \qquad \partial_{\hat t} \Vh=\partial_{\hat t} \Kh=0.
\end{equation}

In the first two cases $f$ is fixed to be power-law. Note that, as with the $f(\phi)=\phi^2$ case discussed in the Introduction, this means that $\tilde{\phi}$ should be restricted to be positive. Although it is not clear how one implements such a constraint in practice in the functional integral \eqref{equ:pathint-conffact}, the reader can check that the constraint has no influence on the steps taken to derive either the flow equation or the msWI both at the exact level and in terms of the derivative expansion in sec. \ref{sec:BeyondLPA}. Therefore we arrive at the same final equations but with restrictions on the range, in particular $\ph$ should be restricted to be positive in the power-law case. Such restrictions in principle affect solutions of these equations \cite{DietzMorris:2015-2} but will not play a significant r\^ole in this paper.

Summarising, since the reduced equations are independent of the form of $f$, clearly $\hat k$-fixed points are also not affected by the form of $f$. On the other hand the form of $f$ plays a crucial r\^ole for flows in the background-dependent scale $k$, and in particular for flows that satisfy the msWI (as they must given the form of the partition function).  If $\eta\ne0$ and $f$ is not power-law (for example $f$ is of exponential form) then $k$-fixed points are actually forbidden by the required background dependence at finite $k$. If $f$ is power-law but not of the prescribed form:
\be
\label{good-pow}
f(\chi)\propto\chi^{2d_f/\eta}\,,
\ee
then $k$-fixed points are not forbidden but do not coincide with $\hat{k}$-fixed points. Finally if $f$ takes the above prescribed form, or $\eta=0$, then the two notions of fixed points coincide, \viz \eqref{coincide}. In the next section we will gain an understanding of these results in terms of scaling dimensions and renormalization.

\section{Interpreting the equations and their properties}
\label{sec:dimensions}

In the Introduction we already briefly touched on a number of subtle physical issues that lie hidden below the equations. In this section these will be fully explored. In particular this allows us to {explain key results we uncovered in the previous section}.

\subsection{Scale dependence and renormalization}
\label{sec:renormalization}

In the functional RG approach and in the effective action, we can usually ignore the distinction between bare and renormalised quantities, providing we are concerned only with fixed points and providing we identify dimensions of quantities with their scaling dimensions (see \eg the lectures \cite{Morris:1998}). This is a consequence of the fact that nothing runs and fundamentally there is no scale in the problem, so setting the scale in the effective action  to $k$ (the infrared scale introduced by hand) and using this to express all dimensionful quantities in dimensionless terms, effectively takes care of the renormalization and no further renormalization is necessary. Renormalization is then only needed if we move away from the fixed point, thus introducing a mass-scale and running couplings, see \eg \cite{Morris:1996xq}. 

However in the present case we must pay attention to the fact that the background metric ${\bar g}_{\mu\nu}$ receives its definition in the ultraviolet, as part of the bare action and functional integral. This includes for example its r\^ole in the source term and insertion of the infrared cutoff \cf eqns. (\ref{equ:pathint-conffact}-\ref{equ:source}). Through the derivation (see secs. \ref{sec:flowequ-derivation} and \ref{sec:sWI-derivation}), it is this {bare} ${\bar g}_{\mu\nu}$ that then appears explicitly in the flow equation \eqref{equ:flowGamma}, msWI \eqref{equ:sWiGamma}, and ensuing equations, culminating in \eqref{equ:sysV-resc} and \eqref{equ:sysK-resc}, and thus also for consistency in the effective action \eqref{eff-ac-intro}.

Since ${\bar g}_{\mu\nu}$ is replaced by a composite operator through \eqref{equ:confpar}, this matters, as we see when we consider what dimension $d_f$ we should assign to $f$. As already discussed in the Introduction, if we regard the reference metric $\hat g_{\mu\nu}$ as on the same footing as the quantum metric $\tgmn$ and background metric $\bgmn$, then we must set $d_f=0$ as is clear from \eqref{equ:confpar}. This works fine if also the dimension of ${\phi}$ is $[\phi]=0$, equivalently $\eta=0$, since then $f(\phi)$ can be any function of the field and still be consistent with dimensions. Indeed from sec. \ref{sec:fpmap} we saw that in this case, $k$-fixed points are allowed and furthermore coincide with the background independent notion of $\kh$-fixed points.\footnote{We saw this holds also for $d_f\ne0$. We comment on this case later.}

However as already mentioned, we have to allow for a non-vanishing anomalous dimension of the conformal fluctuation field $[\vp]=\eta/2$. This is to ensure that the kinetic term for $\vp$ remains finite in the RG evolution. At an operational level, one usually further requires that it remain canonically normalised \ie  $K(0,\chi)=1$. If we want to preserve this normalisation away from a fixed point, we also require $\eta$ to run with scale.\footnote{or more precisely depend on couplings that run with scale} But since constructing a continuum limit requires that the RG trajectory emanate from a fixed point, we can for a local treatment take  $\eta$ to have a fixed value. (This discussion parallels the treatment of fixed points in scalar field theory \cite{Morris:1994ie,Morris:1994jc} which can then be generalised to incorporate a running anomalous dimension for  the full renormalised trajectory\cite{Morris:1996xq}.) Had we been working at $\mathcal{O}(\partial^2)$ or higher with the background field, we would need a similar normalisation for its kinetic term\footnote{taking into account also the mixing term $\sim\partial_\mu\chi\partial_\mu\vp$} and thus an anomalous dimension for $\chi$ which in general would not be the same as $\eta$, reflecting the difference in $\tilde\vp$--$\chi$ interactions versus $\tilde\vp$ self-interactions that are forced by the breaking of split symmetry. Since we treat $\chi$ at the LPA level, its anomalous dimension is in general not determined (see the discussion in ref. \cite{Bervillier:2013}) however  in our case for the left hand sides of the broken split Ward identities \eqref{equ:sWIV-resc} and \eqref{equ:sWIK-resc} to be consistent in terms of dimensions, the background field has to also have $[\chi]=\eta/2$. Thus working at the LPA level for $\chi$, the two anomalous dimensions are forced to be equal ultimately by the fact that the bare action in the functional integral \eqref{equ:pathint-conffact} depends only on the total quantum conformal factor field $\phi=\chi + \tilde \vp$.

Now we see that in order to maintain $d_f=0$, $f$ necessarily contains a dimensionful parameter. Given that $f(\phi)$ is defined by its introduction in the ultraviolet, at a fixed point this parameter can only be $k_0$, the overall ultraviolet cutoff scale that regularises the functional integral, \cf \eqref{equ:pathint-conffact}. In this way we see that $f$ is really a function of the dimensionless combination $\phi/k_0^{\eta/2}$. At a fixed point there are no other dimensionful parameters we could use. (Na\"\i vely we might be tempted to try and escape this conclusion by using $k$ inside $f(\phi)$ instead, but this makes no sense for an ultraviolet quantity,  as well as causing the derivation of the flow equation \eqref{equ:flowGamma}, as given in sec. \ref{sec:flowequ-derivation}, to break down completely.)

Similarly, if $f$ is taken to have a non-zero dimension $d_f\ne0$ then at a fixed point the dimensionful part must be given by the appropriate power of $k_0$. Putting all this together we see that in reality
\be
\label{F}
f(\phi) = k_0^{d_f} {F}\left({\phi\over k_0^{\eta/2}}\right)\,,
\ee
for some dimensionless bare function $F$.
This has two profound consequences. 

Firstly, using $k$ to make a dimensionless version of the parametrisation of the background conformal factor field
\be 
\label{f-bar}
{\bar f}(\cb) = F\left(\cb\ex{\frac{\eta}{2}(t-t_0)}\right) \ex{d_f(t_0-t)}\,,
\ee 
where $t_0 = \ln(k_0/\mu)$,  we see that  generally such insertions run with scale $t$ in such a way that the $t$ dependence cannot be eliminated, unless at the same time we eliminate $f$. In particular it appears in this form in the effective action \eqref{equ:ansatzGamma}. Since in this original background-dependent description we cannot eliminate $f$ because its presence is required by (remnant) diffeomorphism invariance \eqref{remnant}, we now understand why we found in sec. \ref{sec:fpmap} that $k$-fixed points are in general forbidden. 

Secondly, since $k_0$ is to be sent to infinity in order to form the continuum limit, this means that everything that involves $f$, including the flow equation \eqref{equ:flowGamma}, msWI \eqref{equ:sWiGamma} and effective action \eqref{eff-ac-intro} are not yet well defined. $f$, $V$ and $K$, $\vp$ and $\chi$, must all be regarded as bare quantities. We must search for a way to rewrite these expressions as renormalised quantities, \ie such that they can be declared finite and independent of $k_0$ (equivalently $t_0$). But this is precisely what is achieved by the transition to background-independent variables $\tilde{V}$, $\tilde{K}$ and $\tilde{k}$ in \eqref{equ:chvars}! Indeed in terms of these we are left with the reduced equations \eqref{equ:sys-red} in which all dependence on $f$ has disappeared, and thus we are free to declare the tilde quantities to be finite. If we wish we can now define renormalised versions of the original variables via an equation of the same form as \eqref{equ:chvars}:
\begin{equation}
 \label{equ:ren-vars}
 V_R(k,\vp_R,\chi_R) = f_R(\chi_R)^{-\frac{d}{2}}\tilde V(\tilde k,\phi), \ K_R(k,\vp_R,\chi_R) = f_R(\chi_R)^{-\frac{d}{2}+1}\tilde K(\tilde k,\phi), \ \tilde k = k f_R(\chi_R)^\frac{1}{\alpha}\,,
\end{equation}
where $\phi=\vp_R+\chi_R$. Since the transformation is of the same form,  it is guaranteed that the reduced equations \eqref{equ:sys-red} are then equivalent to the original system \eqref{equ:sysV-resc} and \eqref{equ:sysK-resc} with all bare quantities replaced by their renormalised quantities. For the renormalised conformal factor parametrisation, by dimensions and finiteness 
\be
\label{F_R}
f_R(\phi) = \mu^{d_f} F_R\left({\phi\over \mu^{\eta/2}}\right)\,,
\ee
where $\mu$ is the arbitrary finite physical scale introduced in \eqref{time}. But note that, apart from being defined over the same range, the dimensionless renormalised function $F_R$ need not in fact bear any relation to the bare function $F$. Comparing \eqref{equ:ren-vars} and \eqref{equ:chvars} sets up direct relations (generally divergent in the limit $k_0\to\infty$) between renormalised and bare quantities. Thus comparing the two formulae for $\tilde k$ gives $f_R(\chi_R) =f(\chi)$ which, through \eqref{F_R} and \eqref{F}, gives the relationship between bare $\chi$ and renormalised $\chi_R$:
\be
{F}\left({\chi\over k_0^{\eta/2}}\right) = \left(\frac{\mu}{k_0}\right)^{d_f} F_R\left({\chi_R\over \mu^{\eta/2}}\right)\,,
\ee
 and thus through $\phi$ which is common to both, the relationship between $\vp$ and $\vp_R$. Finally the first two equations in \eqref{equ:ren-vars} and \eqref{equ:chvars} then set up the relationship between bare and renormalised potential and kinetic terms.

This is the case for general functions $F$ and $F_R$. We have already highlighted one special case where $\eta=d_f=0$. From \eqref{F}, we confirm that in this case $f$ does not actually depend on $k_0$. Setting $F=F_R$, no renormalization is necessary. Another similar special case where no renormalization is necessary is the power-law \eqref{good-pow} since then dimensions $d_f$ and $[\phi]$ again match, and thus again $k_0$ drops out so that $f(\phi) = F(\phi)$. In these cases we see also that no running is induced by background dependence either, since it also follows from \eqref{f-bar} that ${\bar f}(\cb) = F(\cb)$, \ie is $t$-independent. It is therefore not surprising that these two cases were found in sec. \ref{sec:fpmap} to result in the two definitions of fixed points coinciding.

The remaining possibilities for $f$ and $d_f$ involve dimensional assignments that imply the presence of a bare mass scale, $k_0$. Either  $f(\phi)$ is general and $\eta=0$ but $d_f\ne0$,  or
$f(\phi)\propto\phi^\gamma $ is power-law but not of the prescribed form \eqref{good-pow}.  In both cases we saw in sec. \ref{sec:fpmap} that $k$-fixed points could exist for ${\bar K}$ and $\bar{V}$. However in the first case they coincided with $\kh$-fixed points, whereas in the latter case they did not. Although these conclusions are intruiging, it is not really correct to regard either case as resulting in fixed points with respect to $k$ since in both cases the effective action still evolves with $t$. To see this note that, since the dimensions no longer match up, ${\bar f}(\cb)$ runs with $t$, although for these two cases now in a purely multiplicative way: 
\be
{\bar f}(\cb) = F\left(\cb\right)\ex{\left(\frac{\eta\gamma}{2}-d_f\right)(t-t_0)}
\ee
(with $\eta=0$ in the first case).
Therefore after appropriate multiplicative redefinitions the $t$ dependence could be eliminated. This is of course achieved in the scaled background-independent variables by \eqref{equ:dimless}. The mismatch in dimensions means that it is not achieved in the scaled original variables \eqref{Vh} and \eqref{Kh}. Therefore at $k$-fixed points, although ${\bar K}$ and $\bar{V}$ do not evolve,  the effective action \eqref{equ:ansatzGamma} continues to evolve with $t$ as we claimed. Furthermore in both cases in physical (unscaled) units we have from \eqref{F} that $f(\chi)=F(\chi) \,k_0^{d_f-\eta\gamma/2}$, so the effective action \eqref{equ:ansatzGamma} still depends on $k_0$ and thus still needs renormalization. (This renormalization proceeds straightforwardly from \eqref{equ:ren-vars}, with both $V$ and $K$ renormalising  multiplicatively and from $F(\chi) = (\mu/k_0)^{d_f-\eta\gamma/2} F(\chi_R)$, being multiplicative for $\chi$ also in the power-law case.)

\subsection{Symmetries and dimensions}
\label{sec:symmetries}

As touched on in the Introduction, having a consistent set of dimensional assignments is equivalent to a statement of scale invariance in \emph{theory space}. Thus for example, the conventional choice in quantum field theory on Minkowski space is to provide the coordinates with a mass dimension $[x]_E=-1$, momentum with a mass dimension $[p]_E=1$ and consistently assign such dimensions to all quantities $Q$ in the theory, so that under the scaling $Q\mapsto \lambda^{[Q]_E} Q$, all the equations remain invariant (and thus in particular actions, flow equations, broken Ward identities \etc remain invariant). It is a symmetry of ``theory space'', mapping from one theory to another, rather than a symmetry of any particular theory, because anything with a non-zero mass dimension including massive couplings and masses $m$, will also get scaled in the process, thus \eg $[m]_E=+1 \implies m\mapsto \lambda m$. Nevertheless the existence of a consistent set of dimensional assignments is an important constraint (as any physics undergraduate ought to know). 

As argued in the Introduction, a natural choice is then to set $[f]_E=0$. Identifying these assignments with the so-called ``engineering dimensions'', 
in other words not taking into account anomalous scaling, we also set $[\phi]_E=0$. Then from \eg \eqref{equ:ansatzGamma}, we read off $[V]_E = d$ and $[K]_E= d-2$.

However in a gravitational theory and in particular in the present conformally reduced theory, there is more than one way to consistently assign dimensions. As we have already noted in the Introduction, the `diffeomorphism' invariance \eqref{remnant}, since it is a consistently realised scaling symmetry (in theory space rather than a particular theory except for the exponential parametrisation), it  can also be regarded as a consistent set of dimensional assignments $[x]_D = -1$, $[f]_D=2$, extended to $[p]_D=1$ in equation \eqref{p-diffeo}. All other quantities have zero dimension: $[\phi]_D=[V]_D=[K]_D=0$. 

The existence of a consistent set of anomalous scaling dimensions imply a further set of consistent dimensional assignments which can be taken just to be the multiple of $\eta$. Thus $[\phi]_A=1/2$, $[K]_A=-1$, whilst $[x]_A=[p]_A=[V]_A=0$. (See \eg  \eqref{equ:dimless}, or \eqref{Vh} and \eqref{Kh}.)

Since these are three different scaling symmetries, we can then assign dimensions according to any `diagonal' subgroup of our choosing, \ie without loss of generality define for all quantities $Q\mapsto \lambda^{[Q]} Q$, where $[\ ]= a\, [\ ]_A + b\,[\ ]_D + c\, [\ ]_E$ for some real numbers $a,b,c$. These dimensional assignments thus form a vector space, spanned by $[\ ]_A$, $[\ ]_D$ and $[\ ]_E$. For our system of equations the vector space is in fact three dimensional, \ie this is the maximum number of linearly independent dimension assignments.

Since $\phi=\chi+\vp$, the dimensions of the components are dictated from $[\phi]$, as well as that of the quantum variable: $[\chi]=[\vp]=[\tilde{\vp}]=[\phi]$. Through \eqref{equ:cutoff-action} the dimension of the cutoff operator $R$ is then dictated. Finally through the choice \eqref{equ:cutoff} with power-law \eqref{equ:cutoff-pwrlw}, this determines also a consistent choice of dimension for $k$. For the three linearly independent choices discussed, the results are displayed in table \ref{table-dimensions}. The resulting dimensional assignments for $k$ may look funny but we emphasise that they are fixed self-consistently with the other assignments in order to be a symmetry of all the equations. The reader can check that with these assignments, all equations in the paper obey these three scaling symmetries, in particular also the flow equations and broken split Ward identities in their various forms.

\begin{table}
\begin{center}
\begin{tabular}{|c||c|c|c|}
\hline 
    $I=$   & $A$ & $D$ & $E$ \\ \hline\hline
$[p]_I$    & 0 & 1 & 1 \\ \hline
$[x]_I$    & 0 &$-1$ &$-1$ \\ \hline 
$[f]_I$    & 0 & 2 & 0 \\ \hline
$[\phi]_I$ & $\half$ & 0 & 0 \\ \hline
$[V]_I$    & 0 & 0 & $d$ \\ \hline
$[K]_I$    & $-1$& 0 & $d-2$ \\ \hline
$[R]_I$    & $-1$& 0 & $d$ \\ \hline
$[k]_I$    & $-\frac{2}{(d+2n)\alpha}$ & 0 & $\frac{2}{\alpha}$ \\ \hline
\end{tabular}
\end{center}
\caption{\label{table-dimensions} Consistent `Anomalous', `Diffeomorphism', and `Engineering' dimension assignments. Recall that $\alpha$ is given by eqn. \eqref{eq:alpha}.}
\end{table}

A number of other scaling symmetries equivalently dimensional assignments are at play, but they are all linear combinations of those in table \ref{table-dimensions}. For example, na\"\i vely scaling dimensions are given by 
\be
[\ ]_{\rm scaling} = [\ ]_E + \eta\, [\ ]_A\,.
\ee

There is an `unphysical scaling symmetry' determined by the power-law cutoff \eqref{equ:cutoff-pwrlw}   \cite{Morris:1994ie}:
\be 
[\ ]_U = [\ ]_E + (d+2n)\, [\ ]_A\,,
\ee
under which the cutoff-scale is invariant: $[k]_U=0$. It is this scaling symmetry that allows the anomalous dimension to be quantised in \eg scalar field theory \cite{Morris:1994ie,Morris:1994jc,Morris:1998,Bervillier:2013}.

One can also recognise a `Weyl dimension' where dimensions are chosen in order to balance the powers of $f^{1/2}$:
\be 
[\ ]_W =    [\ ]_E -  [\ ]_D\,,
\ee
or equivalently
according to the scaling of the fields under (global) Weyl transformations $f\mapsto \lambda^{-2}f$, leaving coordinates invariant. Indeed under these we see that $[x]_W=[p]_W=0$, also $[\phi]_W=0$, whereas $[f]_W=-2$, $[V]_W= d$ and $[K]_W=d-2$. In the framework of quantum gravity, where we would like to consider arbitrary diffeomorphisms as functions of the coordinates, this is arguably a more natural choice of dimension assignment. Let us note this symmetry  ensures that the theory does not depend on Newton's constant $G$ and the cosmological constant $\Lambda$ separately but only on a combination, in an analogous sense to that argued in ref. \cite{Hamber:2013rb}. Indeed, following our remarks below \eqref{eff-ac-intro}, let us define according to some chosen background $\chi$, $K(0,\chi)= 1/(8\pi G)$ and $V(0,\chi) \equiv V_0 = \Lambda/(16\pi G)$.\footnote{$V_0$ is analogous to $\lambda_0$ in ref. \cite{Hamber:2013rb}. Strictly we should here be using renormalised variables as described in the previous section, or assume a parametrisation for $f$ that does not require renormalization.} Under the Weyl scaling, we can absorb $G$ into $f$ and $V$, so that the normalisation conditions only define the combination $G V_0^{1-2/d}$:
\be
V(0,\chi') = \half (8\pi G)^\frac{2}{d-2}\Lambda  = \left( 8\pi G\, V_0^{1-2/d} \right)^{d\over d-2}\,,
\ee 
where the change in $f$ is compensated by a change in background $\chi\mapsto\chi'$. 

Finally, we can also justify the dimension assignments we have been assuming all along, since these are also a linear combination of the dimensions in table \ref{table-dimensions}:
\be 
[\ ] = \left(1-\frac{d_f}{2}\right) [\ ]_E + \frac{d_f}{2}\, [\ ]_D + \eta\, [\ ]_A\,.
\ee
Thus from the table, $[p]=1$, $[x]=-1$, $[\phi]=\eta/2$, $[V]=d-d\,d_f/2$, $[K]=d-2-\eta-(d/2-1)\,d_f$, $[R]=d-\eta-d\,d_f/2$ and in particular $[f] = d_f$ and $[k]=1$.

Once  background independent variables are adopted and dimensionless combinations chosen as in \eqref{equ:dimless}, the 
scale ($\kh$)  no longer appears explicitly in the equations \eqref{equ:sys-red-final}. The action of the scaling symmetries on the remaining variables $\Vh$, $\Kh$, $\ph$ and $\hat p$ is such that $[\ ]_D = [\ ]_E = -\eta\, [\ ]_A$, \ie they are all proportional to one another and only one scaling symmetry survives. The simplest normalisation is provided by the unphysical scaling symmetry: 
\be
[\hat p]_U=1\,,\quad [\Vh]_U =d\,,\quad[\Kh]_U = -2(n+1)\,,\quad[\ph]_U =(d+2n)/2\,.
\ee

\section{Comparisons}
\subsection{Comparison to scalar field theory}
\label{sec:comp-sft}
The flow equation for the effective action $\Gamma[\phi]$ of a standard scalar field $\phi$ is analogous to \eqref{equ:flowGamma} and reads
\be
\label{equ:FRGE}
\partial_t \Gamma = \frac{1}{2}\,\mathrm{Tr}\left[\frac{\delta^2\Gamma}
				  {\delta \phi \delta \phi}+ R_k\right]^{-1} \partial_t R_k,
\ee
where we have left the $k$-dependence of the effective action implicit. To compare the present setup to standard scalar field theory, we note that if we absorb all factors containing the parametrisation function $f$ in \eqref{equ:ansatzGamma} into $K$ and $V$ and let $V \mapsto -V$ the result is
\begin{equation} \label{Gamma-sf}
\Gamma \mapsto -\Gamma_{\mathrm{sf}} = - \int d^dx \left(\frac{1}{2}K\left(\partial_\mu \vp\right)^2
		      +V\right),
\end{equation}
where $\Gamma_\mathrm{sf}$ is the ansatz for the effective action one would write down in the derivative expansion up to $\mathcal{O}\!\left(\partial^2\right)$ of standard scalar field theory. Strictly speaking this is true up to the fact that here $K$ and $V$ also depend on the background field, but as far as the flow equation \eqref{equ:FRGE} is concerned the background field just appears as a parameter and does not affect the conclusions we will come to here. There is also the additional difference given by the choice of dimensions as reflected in \eqref{Vh} which will be appropriately taken care of in a moment. Motivated by the overall minus sign included in the cutoff \eqref{equ:cutoff} reflecting the conformal factor sign in quantum gravity, we also make the replacement $R_k \mapsto -R_k$. Taken together these changes convert the flow equation \eqref{equ:FRGE} into
\begin{equation}\label{flow-comp}
\frac{\partial}{\partial t}\Gamma_\mathrm{sf} = -\frac{1}{2} \, \mathrm{Tr} \left[ \left(\Gamma^{(2)}_\mathrm{sf} + R_k\right)^{-1} \frac{\partial}{\partial t} R_k \right],
\end{equation}
which is the flow equation for the standard scalar field theory effective action $\Gamma_\mathrm{sf}$ of \eqref{Gamma-sf} with an additional minus sign on its right hand side.

Let us pause here to see how these alterations can be connected with the use of the msWI that led to the change of variables \eqref{equ:chvars}. As mentioned before, the effect of this change of variables is to convert the ansatz \eqref{equ:ansatzGamma} for the effective action into \eqref{equ:Gamma-hat}, which is precisely of form \eqref{Gamma-sf} if we also replace $V \mapsto -V$.  Moreover, we have seen earlier that the new RG scale $\kh$ defined by \eqref{equ:kbar} and the last equation in \eqref{equ:chvars}, \ie \eqref{khat}, converts the cutoff operator \eqref{equ:cutoff} into the simpler version \eqref{cutoff-kbar}. The $f^{d/2}$ factor in this expression takes care of the measure factor in \eqref{equ:cutoff-action}, leaving a cutoff action that takes the form used in standard scalar field theory up to its overall sign and the replacement $\eta \mapsto d-2+\eta$. Substituting for the anomalous dimension according to this rule is necessary to pass from the dimension of the conformal factor field to the dimension of a standard scalar field. Hence, after implementing the replacement rules 
\begin{equation}\label{replsf}
V \mapsto -V \qquad \text{and} \qquad \eta \mapsto d-2+\eta
\end{equation}
in \eqref{equ:sys-red-final}, we obtain the same flow equations as for standard scalar field theory with the replacements $\Gamma_\mathrm{sf}\mapsto -\Gamma_\mathrm{sf}$ and $R_k \mapsto -R_k$, i.e. the flow \eqref{flow-comp}.

This relation to scalar field theory can be exemplified in $d=3$ dimensions. Applying \eqref{replsf} to the reduced flow \eqref{equ:sys-red-final} and evaluating the integrals reproduces the scalar field theory flow equations of ref. \cite{Morris:1994ie} with an additional minus sign on their right hand sides as in \eqref{flow-comp}.\footnote{When the equations are compared it should be noted that the RG time used in \cite{Morris:1994ie} corresponds to $-\hat t$ in the present context and the right hand sides are rescaled with the factor $1/(2\pi)$ compared to \eqref{equ:sys-red-final}.}

The second replacement rule in \eqref{replsf} simply amounts to a redefinition of a parameter and does not represent a structural difference to scalar field theory. Similarly, at a mathematical level, changing the sign of the potential is only a simple change of variables for the equations \eqref{equ:sys-red-final}. Of course, at a physical level passing from $V$ to $-V$ is not innocuous since it can lead to a potential unbounded below. But as far as the mathematical analysis of the differential equations \eqref{equ:sys-red-final} is concerned, we conclude that the only relevant difference to scalar field theory is a relative minus sign between its left and right hand sides. As noted in ref. \cite{Bonanno:2012dg} this sign nevertheless in principle forbids the existence of well-defined RG flows towards the infrared (instead allowing such flows only towards the ultraviolet) thus undermining the Wilsonian interpretation, with potentially profound consequences.

Note that it is impossible however to eliminate this relative minus sign by further redefinitions whilst leaving all other terms of the same form. In  \eqref{equ:sys-red-final}  the sign of the $\partial_{\hat t}$ is readily changed by reversing the direction of RG flow, and the $\eta/2$ terms are readily changed by redefining $\eta\mapsto-\eta$, but the sign cannot be flipped on the dimension assignments in front of $\Vh$ and $\Kh$ since they had to take these values in order to eliminate $\kh$ and thus $k$ from these equations, \cf \eqref{equ:sys-red}. Such a relative minus sign is therefore meaningful not just for flows but also at fixed points. Indeed we will see in a companion paper that the fixed point solutions are not related to those in scalar field theory \cite{DietzMorris:2015-2}.

\subsection{Comparison with the earlier work}
\label{sec:earlier}

The question arises how these results compare to the study \cite{Manrique:2009uh}, where the RG flow of the conformal factor has been investigated within a bi-field LPA. The ansatz for the effective action in \cite{Manrique:2009uh} reads
\begin{equation}
  \label{equ:Gamma-Reuter}
  \Gamma_k[\phi,\chi] = -\frac{3}{4\pi}\int d^4x\left\{\frac{1}{2G_k}\left(\partial_\mu \phi\right)^2+ \frac{1}{2G_k^B}\left(\partial_\mu \chi\right)^2+ W_k(\phi,\chi)\right\}.
 \end{equation}
It is expressed as a functional of the total conformal factor field $\phi = \chi+\vp$, where $\vp$ and $\chi$ refer to the variables used here in the original flow equations \eqref{equ:flowV} and \eqref{equ:flowK}, and $W(\phi,\chi)$ is a general scale dependent potential. It also contains the scale dependent analogue of Newton's constant $G_k$ for the total field as well as for the background field, $G_k^B$. The reference metric of \eqref{equ:confpar} is also chosen to be $\hat g_{\mu\nu} = \delta_{\mu\nu}$ in \cite{Manrique:2009uh} and we have implemented this in \eqref{equ:Gamma-Reuter} as well as transcribed into the present notation. After deriving the flow equation and the msWI for the potential $W$, it is then found that they cannot in general be satisfied simultaneously, as evidenced in particular by the fact that they cannot be satisfied simultaneously at a non-Gaussian fixed point. Importantly, as in \eqref{equ:cutoff}, the cutoff operator in \cite{Manrique:2009uh} is also implemented with respect to the covariant background field Laplacian. However, a crucial difference between \eqref{equ:Gamma-Reuter} and our ansatz for the effective action \eqref{equ:ansatzGamma} is that the kinetic terms in \eqref{equ:Gamma-Reuter} are not formulated in terms of the background field covariant derivative, nor is the appropriate measure factor containing the background field included in \eqref{equ:Gamma-Reuter}. As a consequence the remnant diffeomorphism invariance \eqref{remnant} and the resulting tight constraints on the structure, are not realised. If the background field is varied, the operator describing high and low momentum modes of the classical conformal factor in \eqref{equ:ansatzGamma} changes with it as dictated by this invariance and as the gravitational setting requires, whereas the high and low momentum modes of the ansatz \eqref{equ:Gamma-Reuter} remain the same. Due to these differences, a direct comparison of results is not possible.

Although we will not pursue the connection,  let us note that the symmetries, in particular Weyl transformations $[\ ]_W$ discussed in subsec. \ref{sec:symmetries}, background independence through shift Ward identities, and maintenance of remnant diffeomorphism invariance \eqref{remnant} through the background covariant cutoff \eqref{equ:cutoff-action}, play a r\^ole in the discussions of Weyl invariance and Weyl invariant regularisation in refs. \cite{Machado:2009ph,Percacci:2011uf}, where however these invariances are imbued with extra meaning. In ref. \cite{Percacci:2011uf} it looks like the Weyl invariant formulation RG equations are non-autonomous, forbidding the existence of fixed points, although this issue is solved in ref. \cite{Pagani:2013fca}.  As above \cite{Manrique:2009uh}, the similarities with our work however seem to be only superficial, and in particular there is no connection to our proof in sec. \ref{sec:fpmap} that $k$-fixed points are forbidden for general parametrisations.

\section{Summary and conclusions}
\label{sec:conclusions}

A physical definition of scale necessarily depends on the metric. As we discussed in the Introduction, since in quantum gravity the metric itself must fluctuate, it is no longer immediately clear what is meant non-perturbatively by such a basic notion in interacting quantum field theory as scale invariance, and more generally scale dependence as expressed through the Renormalization Group (RG).  This issue is therefore certainly important for the asymptotic safety programme. But it is obviously important for any theory of quantum gravity, since if it is to be phenomenologically successful the theory has to allow a large separation of `scales', whatever this is supposed to mean, between the short distances where quantum gravity is believed to be important and the large distances where gravity is known experimentally to behave essentially classically.

Utilising the background field method whereby dependence on scale $k$ is measured through some (general unspecified) background metric $\bgmn$ does not on its own resolve the issue since the notion of scale is then inherently dependent on this $\bgmn$. As argued in general in the Introduction and demonstrated in detail within the conformally reduced quantum gravity model in sec. \ref{sec:dependence}, this means that solutions of the RG flows in general then depend on the choice of background metric. 

Within the conformally reduced gravity model, and following the well understood  route\cite{Pawlowski:2005xe,Litim:2002hj,Bridle:2013sra,Reuter:1997gx,Litim:1998nf,Litim:2002ce,Manrique:2009uh,Manrique:2010mq,Manrique:2010am}, imposing broken split Ward Identities at any finite scale $k$, if everything is done exactly, is then sufficient to ensure that the theory depends only on the total field $\phi=\vp+\chi$ in the limit as $k\to0$, since if the msWI is satisfied at any $k$ it is then satisfied at all scales $k$ where the solution for the effective action makes sense, and thus also in the limit $k\to0$ where the broken Ward Identity turns into the exact Ward Identity.  This limit has however to be taken at fixed finite physical momenta and fields, \ie unscaled, so that these values become ever larger with respect to $k$, and thus the term in braces in \eqref{equ:sWiGamma} vanishes as a consequence of $R_k$ also vanishing in this limit, clearly turning the modified Ward Identity into the exact Ward Identity.

There are two evident problems with this prescription as so formulated however. Firstly since it is practically impossible to solve either the flow equations or the msWI exactly, it is typically no longer guaranteed in practical calculations that the theory depends only on the total field $\phi$ in the limit $k\to0$. The second problem is potentially much more severe and is one of principle. RG properties are realised with a running (thus non-vanishing) $k$ and utilising scaled momenta and fields, and thus the description remains necessarily background dependent. Background independence can at least in principle recovered, but only as $k\to0$, and in unscaled units, where the RG properties will no longer be manifest. 

It would seem that following this strategy, this is the best we can hope for. Both RG properties and background independence can be incorporated, but in a way in which only one of the two properties is manifest in any given description. This however leaves open the possibility that the two properties are actually incompatible at some level. In this paper we have seen that in general this is in fact the situation that arises, at least within the conformally reduced model we investigated. We showed in sec. \ref{sec:fpmap} that for anomalous dimension $\eta\ne0$ and  a general choice of parametrisation $f(\phi)$,  RG fixed points are actually forbidden on any flows that satisfy the msWI. 

It is important to understand the impact that this has. As emphasised in the Introduction, the msWI is not an optional extra. Providing the partition function can be defined, background independence is built in to the original partition function by ensuring dependence only on the total field $\tilde\phi$. This in turn leads to the msWI which thus \emph{must} be satisfied at all points $k$ on the flow if we are to be describing the physical system that was  intended. Fixed points are not guaranteed but certainly a desired centrally important property: without these, again we have no notion of scale invariance. The Wilsonian RG construction of a renormalised trajectory emanating from a fixed point cannot be implemented, and thus according to our current understanding there is no sense in which a continuum limit can be constructed. Therefore the general case, where RG fixed points are forbidden, leads to the worst-case scenario for the conflict between RG notions and background independence mentioned in the first paragraphs above and in the Introduction.

The research reported on in this paper was inspired by the discovery in a scalar field theory model, of a transformation to background-independent variables including a background-independent notion of scale, $\kh$,  which thus automatically ensures that the msWI is then satisfied all along the flow \cite{Bridle:2013sra}. However this transformation was guessed at, using physical intuition.  In appendix \ref{app:scalar}, we showed that the changes of variables discovered in ref. \cite{Bridle:2013sra}, could have been deduced from the structure of the equations, namely the fact that the flow equation and msWI could be combined to give a linear partial differential equation, whereupon the change of variables is found by the method of characteristics. 

In the current case, given that (remnant) diffeomorphism invariance does not allow background dependence to be removed from the cutoff, it is not clear {\it a priori} that a background independent RG description should exist at all. However in sec. \ref{sec:independence} we saw that flow equations and msWIs can be combined to give linear partial differential equations if $\eta=0$ or/and a power-law profile \eqref{equ:cutoff-pwrlw} is chosen. This in turn led to changes of variables which reduce the system of equations to manifestly background-independent flows \eqref{equ:sys-red-final} in a new scale $\kh$ given by \eqref{khat}.  These guarantee that the msWIs and flow equations are simultaneously satisfied, and unify together in one description both background independence, and RG properties such as fixed points, RG trajectories and so on. Therefore $\kh$ defines a background-independent notion of scale, providing an answer to the questions raised in the first paragraphs of the Introduction and Conclusions.

Currently it is not clear whether the cutoff being power-law is then a necessary condition for the existence of background-independent variables in the case $\eta\ne0$. It seems reasonable to conjecture that for a different cutoff form the change of variables from combining the msWIs with the flow equations, still exists, but takes an implicit form such that in principle there also is a set of reduced equations which would lead to qualitatively the same RG structures, as could be argued for in general on the grounds of universality.

We found further convincing evidence that these reduced equations \eqref{equ:sys-red-final}
uncover  a deeper level of description. Firstly, the equations are also manifestly independent of the form of the parametrisation $f(\phi)$ and of its dimension. This means of course that the existence or otherwise of $\kh$-fixed points, their properties, and the structure of $\kh$ flows in theory space, are also independent of these details. They are universal in this sense. In the literature various choices have been advocated, for example $f(\phi) = \exp(2\phi)$ \cite{Machado:2009ph} or $f(\phi)=\phi^2$ \cite{Bonanno:2012dg,Manrique:2009uh}. The existence of the reduced equations demonstrates that it is possible to quantise the conformal factor in a way which is completely independent of such choices (including the possibility that $f$, and thus the metric, actually vanish for some values of $\phi$, and up to topological restrictions on the range of $\phi$ as discussed at the end of sec. \ref{sec:fpmap}). More generally there is advocacy for various parametrisations of the metric, for example \cite{Percacci:2011uf,Nink:2014yya,Falls:2015qga,Percacci:2015wwa} and for the spin connection \cite{Gies:2013noa,Gies:2015cka,Lippoldt:2015cea}. It is tempting to conjecture that in fully fledged gravity there also  exists a deeper analytical description which is manifestly independent of these details. 

Exploiting the fact that solutions of the reduced equations are in one-to-one correspondence with those solutions of the original flow equations that also obey the msWIs, in secs. \ref{sec:fpmap} and \ref{sec:renormalization} we saw that such flows with respect to $k$ are very much affected by the choice of parametrisation $f$. As already reviewed, in the general case fixed points are actually forbidden. However for special cases, in particular if $\eta=0$ or if $f$ takes the prescribed power-law form \eqref{good-pow}, the scaled form of $f$ does not run with $t$ and also we saw that $k$-fixed points are then allowed by the msWIs. Furthermore we saw that in these cases the $k$-fixed points coincide with the $\kh$-fixed points, which of course were there all along since their existence and properties are independent of the choice of $f$. (We also saw that there were special cases where $k$-fixed points are allowed, which either coincide with $\kh$-fixed points [$\eta=0$, $d_f\ne0$] or do not coincide with $\kh$-fixed points [$\eta\ne0$, $f$ power-law but not of form \eqref{good-pow}], but which do not fully correspond to  fixed points since, although the effective potential $V$ and kinetic term $K$ are fixed,  the effective action \eqref{equ:ansatzGamma} still runs with $t$ in these cases because $f$ still runs with $t$.)

The cases where the scaled version of $f(\phi)$ runs with $t$ correspond to cases where $f$ contains a dimensionful parameter. As argued in sec. \ref{sec:dimensions} this dimensionful parameter is set at the overall bare cutoff scale and thus can be taken to be proportional to $k_0$, as introduced below \eqref{equ:pathint-conffact}. If quantum gravity (or rather here the conformally truncated model) is to have a continuum limit, we need to have a renormalised description which is independent of the ratio $k_0/\mu$ as $k_0/\mu\to\infty$. Even if we regard the description as an effective theory only, where we might naturally put $k_0\sim M_{\rm Planck}$ (the Planck mass), in order to achieve an effective continuum description we still require such a description. The background independent variables of the reduced equations automatically provide such a description but are not directly related to physical observables. Again using the one-to-one correspondence we saw in \ref{sec:renormalization} how then to renormalise the original background-dependent variables. In particular we saw that, apart from being defined over the same range, there need not be any direct relation between the bare parametrisation $f$ and renormalised parametrisation $f_R$. The existence of a background independent realisation of the flow equations therefore allowed us to renormalise the equations and thus remove the implicit dependence on the overall cutoff $k_0$. Had we not been able to do so, again there would have been no strict sense in which RG could be formulated in this context since the equations would have remained dependent on  two scales, the bare scale $k_0$ and the infrared scale $k$. It is tempting therefore to conjecture that this holds in general: background independent flow equations necessarily exist if renormalization is possible in such gravitational systems.

We established all these properties in a very general (non-polynomial) truncation defined by keeping all quantities to $\mathcal{O}(\partial^2)$ in the fluctuation field $\vp$, while taking the slow background field limit. The latter amounts to working in the LPA for $\chi$. In sec. \ref{sec:BeyondLPA}, we generalised the algorithmic technique put forward in ref. \cite{Morris:1994ie} for computing the functional trace order by order in the derivative expansion, to both the flow equation and msWI in the current model. As discussed in sec. \ref{sec:evalsWI} some care is needed in evaluating the msWIs in the slow background field limit. By taking the derivative expansion to $\mathcal{O}(\partial^2)$ in the fluctuation field $\vp$, this is already a significant step beyond the scalar field theory model in ref. \cite{Bridle:2013sra}, which was treated only at the LPA level. This step is necessary if we are to treat reliably the case of non-vanishing anomalous dimension $\eta$. We will investigate this along with the fixed point structure in ref. \cite{DietzMorris:2015-2}.
Of course an even more significant step is the treatment of a case where background dependence is forced by the structure of the equations. In close analogy with fully fledged quantum gravity, requiring background dependence in such a way as to respect remnant diffeomorphism invariance \eqref{remnant}, forced the appearance of the background metric \eqref{equ:choiceBackg}, equivalently $f(\chi)$ terms, in the effective action \eqref{equ:ansatzGamma}, the flow equation \eqref{equ:flowGamma-detail} and msWI \eqref{equ:sWI-detail}. The use of the slow field limit for $\chi$ gives this remnant symmetry maximum power in the sense that no space-time derivatives of $f(\chi)$ can appear, and the appearances of $f(\chi)$ are then determined uniquely. Had we worked beyond LPA for $\chi$ we would need local diffeomorphism invariance to constrain the background dependence. At the same time use of the momentum representation would no longer be straightforward, nor would it be necessarily justified to make the tacit assumption (built into the momentum representation) that the space-time manifold is covered by one single patch of infinite extent. 

Staying within the conformally truncated model but going to $\mathcal{O}(\partial^4)$ would allow us to include the effect of the Weyl anomaly in $d=4$ dimensions, see \eg \cite{Machado:2009ph}. In $d=2$ dimensions, going to $\mathcal{O}(\partial^2)$ in the background field $\chi$ would allow us to make contact with Liouville field theory \cite{Distler:1988jt}. The similarity of the reduced equations to scalar field theory as investigated in sec. \ref{sec:comp-sft}, already brings us tantalisingly close, however $\mathcal{O}(\partial^2)$ for $\chi$ also, is needed to address the Weyl anomaly and dependence on background curvature.

From the treatment given here, it still looks like a large step to these cases and indeed to fully fledged quantum gravity. In particular for a full treatment at least in the usual approach we would need to include gauge fixing and ghosts, with the consequences that were discussed in the Introduction. For these reasons, refs. \cite{Branchina:2003ek,Pawlowski:2003sk,Donkin:2012ud,Demmel:2014hla} perhaps furnish a promising direction to take these ideas forward. With the significant increase in complexity however comes also a powerful increase in symmetry. This paper illustrates that symmetries can be equally powerful even if broken by the cutoff, by requiring the appropriate modified Ward Identities to be fully incorporated. Already in the current model, the expression for the background-independent scale \eqref{khat} could not easily be guessed. It seems the nearest to a direct route would have been to deduce it from the assumed scaling dimensions, if one can argue that the cutoff must transform from form \eqref{equ:cutoff} into the form \eqref{cutoff-kbar}. In the current model, multiple scaling symmetries (three dimensions worth) also seem to be important for the existence of a background-independent description (and not just in allowing scale-invariant parametrisations),  since as we saw in sec. \ref{sec:dimensions}, the background-independent variables turn out to be invariants under a two-dimensional subgroup of these symmetries.  Finally, let us emphasise that we were able to discover the background-independent description by following a constructive mathematical --essentially algorithmic-- process. This would seem to provide the strongest hint on how to proceed in these more complex cases.

\section*{Acknowledgments}
It is a pleasure to thank Alfio Bonanno, Astrid Eichhorn and Roberto Percacci for insightful discussions on a number of issues connected to this paper.


\appendix

\section{Solving for background independent variables in scalar field theory}
\label{app:scalar}

Ref. \cite{Bridle:2013sra} reported on an investigation of background dependence in the theory of a single-component real scalar field $\phi$ in $d$ Euclidean space-time dimensions. An infrared cutoff
\begin{equation} \label{equ:cutoffh1}
 R_k\left(-\partial^2,\pb\right) =\left(k^2+\partial^2-h(\pb)\right) \theta \! \left(k^2+\partial^2-h(\pb)\right)\,,
\end{equation}
was incorporated, that depended on an arbitrary function $h$ of a background scalar field $\pb$ 
and RG time $t$. The cutoff breaks the split symmetry that encodes the fact that the fundamental scalar theory depends only on the total classical field $\phi=\pb+\vp$.
Working at the LPA level but retaining separate dependence on the two fields,
 the flow equation for the effective potential $V(\vp,\bp)$ was found to be (in dimensionless variables):
\be
\label{flow2}
\partial_t V - \frac{1}{2}(d-2)\left(\vp \partial_\vp V +\pb \partial_{\pb}V\right) +dV 
 = \frac{(1-h)^{d/2}}{1-h+\partial^2_{\vp}V} \left(1-h-\frac{1}{2}\partial_t h +\frac{1}{4}(d-2)\pb h'\right)\theta(1-h)\,,
 \ee
and the broken split Ward Identity was found to be:
\be
\partial_\vp V-\partial_{\pb} V = \frac{h'}{2} \frac{(1-h)^{d/2}}{1-h+\partial^2_{\vp}V}\,\theta(1-h)\,. \label{sWI-LPA}
\ee
It was then demonstrated that the change of variables
\begin{equation} \label{changevars}
 V = (1-h)^{d/2}\hat V, \qquad \vp = (1-h)^\frac{d-2}{4} \hat \vp -\pb, \qquad t = \hat t -\ln \sqrt{1-h}
\end{equation}
in the region $h(\pb)<1$, allows the two equations to be combined into a single universal flow equation where all background dependence, and dependence on $h$, has disappeared. 

The change of variables was arrived at by inspection, recognising that 
in terms of the running scale $k$, the last equality translates in dimensionful variables  into $\hat k =\sqrt{k^2-h}$, which is precisely the replacement that
is needed in \eqref{equ:cutoffh1} in order to transform this cutoff operator into the standard field independent
cutoff operator where $h=0$ \cite{Bridle:2013sra}. In dimensionless variables, this transformation of the renormalization group scale
$k$ induces corresponding transformations on the field and the potential as expressed in the first two equations
in \eqref{changevars}. 

Here we want to show that this change of variables can be derived directly from the structure of the equations \eqref{flow2} and \eqref{sWI-LPA}. This approach is therefore more powerful and generally applicable,  in particular to the gravitational case as treated in sec. \ref{sec:independence}. 

We note that the parts of \eqref{flow2} and \eqref{sWI-LPA} which are non-linear in $V$ appear only on the right hand sides (as a consequence of the general structure of the exact RG and msWI equations). Again as a consequence of their general structure the non-linear dependence is the same, and thus can be eliminated by combining the equations:
\be
\frac{h'}{2} \partial_t V - \frac{d-2}{4} h' (\vp+\pb) \partial_\vp V +\frac{d}2 h' V +\left(1-h-\frac12 \partial_t h\right) \left(\partial_{\pb} V-\partial_\vp V\right) = 0\,.
\ee
The second and last term already imply some simplification if we use  the total field $\phi=\vp+\pb$ instead of $\vp$:
\be
\label{appendix-linear}
\frac{h'}{2} \partial_t V - \frac{d-2}{4} h' \phi \,\partial_\phi V +\frac{d}2 h' V +\left(1-h-\frac12 \partial_t h\right) \partial_{\pb} V = 0\,.
\ee
Since these partial differential equations are linear in $V$, they can be solved by the method of characteristics. From $(\partial_tV,\partial_{\pb} V,\partial_{\phi} V,-1)\cdot (\delta t,\delta \pb,\delta \phi,\delta V)=0$, we identify the normal to the solution surface
and thus from \eqref{appendix-linear} also the vector field that generates characteristic curves 
in the surface depending on some auxiliary parameter $s$:
\begin{subequations}
\label{characteristics}
\begin{align}
\frac{dt}{ds} &= \frac{h'}2\,,\label{c1}\\
\frac{d\pb}{ds} &= 1-h-\frac12\partial_th\,,\label{c2}\\
\frac{d\phi}{ds} &= -\frac{d-2}4\phi\,h'\,,\label{c3}\\
\frac{dV}{ds} &= -\frac{d}2h'\,.\label{c4}
\end{align}
\end{subequations}
Combining \eqref{c1} and \eqref{c2} 
we can regard $\pb$ as a function of $t$ such that
\be
\frac{h'}2 \frac{d\pb}{dt}+\frac12\partial_th = 1-h\,,
\ee
and thus 
\be
\label{appendix-t}
\ln \sqrt{1-h} = -t +\hat t\,,
\ee
where we have introduced the integration constant $\hat t$. Combining \eqref{c1} and \eqref{c4} shows that $V\exp dt$ is an integration constant, or equivalently from \eqref{appendix-t},
\be
\label{appendix-V}
\hat V =  (1-h)^{-d/2} V \,,
\ee
where $\hat V$ is the integration constant. Finally, combining \eqref{c3} and \eqref{c4}, and using \eqref{appendix-V}, we have 
\be
\label{appendix-phi}
\phi = (1-h)^{d-2\over4}\hat\phi \,,
\ee
in terms of some integration constant $\hat\phi$. We see that equations (\ref{appendix-t}--\ref{appendix-phi}) reproduce the desired change of variables \eqref{changevars}.


\bibliographystyle{hunsrt}
\bibliography{refs}

\end{document}